\documentclass[conference]{IEEEtran}
    
\def\BibTeX{{\rm B\kern-.05em{\sc i\kern-.025em b}\kern-.08em
    T\kern-.1667em\lower.7ex\hbox{E}\kern-.125emX}}

\usepackage{algorithmic}
\usepackage{graphicx}
\usepackage{textcomp}
\usepackage{makecell}
\usepackage{booktabs}
\usepackage{multirow}
\usepackage{xcolor}
\usepackage{framed}
\usepackage{url}
\colorlet{shadecolor}{lightgray}

\usepackage[most]{tcolorbox}

\colorlet{barcolour}{red}
{\endMakeFramed}

\tcbset{textmarker/.style={%
        enhanced,
        parbox=false,boxrule=0mm,boxsep=0mm,arc=0mm,
        outer arc=0mm,left=0mm,right=0mm,top=7pt,bottom=7pt,
        toptitle=1mm,bottomtitle=1mm,oversize}}

\newtcolorbox{noteBox}{textmarker,
    colback=gray!10!white}

\newcolumntype{P}[1]{>{\centering\arraybackslash}p{#1}}
\newcolumntype{M}[1]{>{\centering\arraybackslash}m{#1}}

\definecolor{r1}{RGB}{87,114,158}
\definecolor{r2}{RGB}{204,137,99}
\definecolor{r3}{RGB}{93,157,107}
\definecolor{r4}{RGB}{196,78,82}
\definecolor{r5}{RGB}{129,114,180}
\definecolor{r6}{RGB}{147,120,96}
\usepackage{marginnote}
\usepackage{soul} 
\usepackage{hhline}
\definecolor{lightyellow}{RGB}{250, 250, 180}
\definecolor{HLYELLOW}{RGB}{240, 127, 0}
\definecolor{hlyellow}{RGB}{240, 127, 0}
\sethlcolor{lightyellow}
\definecolor{lightcyan}{RGB}{160,255,255}
\usepackage{xparse}

\usepackage{mdframed}
\global\mdfdefinestyle{review}{%
linecolor=lightyellow,linewidth=3pt,%
leftmargin=0cm,rightmargin=0cm,%
skipabove=0cm,skipbelow=0cm,%
innerrightmargin=0cm,innerleftmargin=0cm,%
innerbottommargin=0cm,innertopmargin=0cm,%
backgroundcolor=lightyellow
}
\global\mdfdefinestyle{reviewtext}{%
linecolor=lightyellow,linewidth=0pt,%
leftmargin=0cm,rightmargin=0cm,%
skipabove=0.1cm,skipbelow=0.1cm,%
innerrightmargin=0cm,innerleftmargin=0cm,%
innerbottommargin=0cm,innertopmargin=0cm,%
backgroundcolor=lightyellow
}

\begin{document}

\DeclareDocumentCommand\review{m g g}{%
    {\IfNoValueF {#2}{%
    \IfNoValueF {#3}{%
    {\marginnote{\tcbox[colframe=#3,on line,boxsep=0pt,left=3pt,right=3pt,top=3pt,bottom=3pt,colback=#3]{\normalfont \textbf{{\normalsize{\color{white}#2}}}}}%
    }%
    }%
    \IfNoValueT {#3}{%
    {\marginnote{\normalfont \textbf{\normalsize{#2}}}%
    }%
    }%
    }%
    \hl{#1}%
    }%
}

\DeclareDocumentCommand\reviewall{m g}{%
    {\IfNoValueF {#2}{%
    {\marginnote{\tcbox[colframe=black,on line,boxsep=0pt,left=3pt,right=3pt,top=3pt,bottom=3pt,colback=white]{\normalfont \textbf{{\normalsize{\color{black}#2}}}}}%
    }%
    }%
    \hl{#1}%
    }%
}

\title{Noise in the Clouds: Influence of Network Performance Variability on Application Scalability}


\author{\IEEEauthorblockN{Daniele De Sensi, Tiziano De Matteis, Konstantin Taranov,\\
                          Salvatore Di Girolamo, Tobias Rahn, and Torsten Hoefler}
        \IEEEauthorblockA{\\
                Department of Computer Science, ETH Zurich, Switzerland\\
                \{first-name.last-name\}@inf.ethz.ch}}

\maketitle
\begin{abstract}
Cloud computing represents an appealing opportunity for cost-effective deployment of HPC workloads on the best-fitting hardware. However, although cloud and on-premise HPC systems offer similar computational resources, their network architecture and performance may differ significantly. For example, these systems use fundamentally different network transport and routing protocols, which may introduce \textit{network noise} that can eventually limit the application scaling. This work analyzes network performance, scalability, and cost of running HPC workloads on cloud systems. First, we consider latency, bandwidth, and collective communication patterns in detailed small-scale measurements, and then we simulate network performance at a larger scale. We validate our approach on four popular cloud providers and three on-premise HPC systems, showing that network (and also OS) noise can significantly impact performance and cost both at small and large scale.
\end{abstract}


\section{Introduction}
Due to flexibility and cost-effectiveness, running HPC applications in the cloud has become an appealing solution and a potential alternative to on-premise systems~\cite{2203.02544, parashar13}. Scientific applications from different domains already run on the cloud, including multiphysics simulations~\cite{PENAMONFERRER202114, 9721335} and biomedical applications~\cite{10.1007/978-3-030-69984-0-53,  10.1371/journal.pcbi.1006144}. 

One of the main advantages of cloud computing is the possibility to run an application on the most appropriate computational resources in a cost-effective way. Instances that can be deployed in the cloud come with a wide variety of architectural characteristics in terms of memory, CPUs, accelerators, and network bandwidth.
On the CPU side, it is possible to select between different processors, with different numbers of cores, clock frequency, and architecture, ranging from commercial off-the-shelf Intel and AMD processors to custom ARM processors like the ARM Graviton processor deployed by AWS~\cite{graviton}. Cloud providers also offer a wide choice of accelerators that includes different types and generations of GPUs~\cite{awsgpu}, TPUs (\textit{Tensor Processing Units})~\cite{gcptpu}, and FPGAs~\cite{azurefpga}. Similarly, different instances provide different network bandwidths. Users can deploy instances with 100 Gb/s networks on most major cloud providers and, in some cases, even 200 Gb/s and 400 Gb/s instances (on Azure and AWS, respectively).
Finally, cloud vendors frequently deploy new hardware, differently from on-premise HPC systems, where compute resources have life-cycles spanning multiple years. 

However, all this flexibility comes at a cost. Although we can expect minor differences in the compute performance between an HPC instance in the cloud and an equivalent server in an on-premise HPC system~\cite{zhai11, guidi21}, the network performance can significantly differ. Indeed, in some cases, the network connecting those instances in the cloud significantly differs from a traditional HPC network. For example, packets might be routed using ECMP~\cite{ecmp,srd,googlenet} in a congestion-oblivious way, and can thus experience a higher latency if multiple network flows are mapped on the same paths~\cite{10.1145/2535372.2535375,conga,drill}. On the contrary, HPC systems often deploy adaptive routing to react more promptly to congestion in the network~\cite{slingshot,8961159}. Also, differently from most HPC systems, some providers do not use \textit{Remote Direct Memory Access} (RDMA), or run instances on tapered networks~\cite{googlenet}.
All these factors can contribute to increase network latency, decrease network bandwidth, and increase \textit{network noise}~\cite{10.1145/3295500.3356196,10.1145/3126908.3126926,8665797,7877142,hoefler-collnetnoise} (i.e., performance variability induced by the use of the network). This limits the scalability and tampers cost-effectiveness. Although HPC applications can scale up to 42 million cores~\cite{10.1145/3458817.3487399} on on-premise HPC systems, it is still not clear how far HPC applications could scale on the cloud. Traditionally, cloud environments have been considered a good match for loosely coupled or embarrassingly parallel workloads, but network performance has been seen as one of the main bottlenecks preventing their adoption for tightly coupled computations~\cite{jackson10, zhai11,magellan-report11, marathe13, chang18, netto18-survey}.

Assessing the network performance and the impact of noise on scalability is even more relevant if we consider that the gap between compute and network performance increases. For example, \textit{from 2010 to 2018, the computational throughput of the Top 500 HPC systems~\cite{top500} increased by 65x, while the off-node communication bandwidth only increased by 4.8x~\cite{8425191,10.1145/3295500.3356145}}. 
Thus we expect, in the future, network performance to be even more relevant for HPC applications running on the cloud. 

\begin{table*}[h!]
\centering
\footnotesize
\caption{Analyzed systems: for each of them we detail the CPU, Memory and network characteristics. \textit{C} indicates the number of physical cores. Instance costs as referred to the July, 18 2022 for the East US availability zone.}
\label{tab:systems}
\resizebox{\textwidth}{!}{
\begin{tabular}{p{0.5cm} c c c c c c c c c c} 
 & \textsc{System} &  \thead{\textsc{Instance}\\\textsc{Type}} & \textsc{CPU} & \textsc{Memory}   & \thead{\textsc{Per Hour}\\\textsc{Instance Cost}\\\textsc{(Committed)}} & \thead{\textsc{Per Hour}\\\textsc{Instance Cost}\\\textsc{(on-demand)}} & \textsc{Bandwidth}& \textsc{Network} & \textsc{Routing} & \thead{\textsc{Transport}\\\thead{\textsc{Protocol}}} \\
 \midrule
 \multirow{3}{*}[-4.5ex]{\hspace{2ex}\rotatebox[origin=c]{90}{\textsc{AWS}}}   & Normal & c5.18xlarge & \thead{2x18C\\Intel Xeon Platinum\\8124M @ 3GHz}  & 144 GB & 1.34 USD & 3.06 USD & 25 Gb/s &  \thead{Fat Tree~\cite{srd}} & ECMP~\cite{srd} & SRD~\cite{srd} \\ 
                        & HPC (Metal) & c5n.metal & \thead{2x18C\\ Intel Xeon Platinum\\8124M @ 3GHz} & 192 GB & 1.475 USD & 3.88 USD & 100 Gb/s & \thead{Fat Tree~\cite{srd}} & ECMP~\cite{srd} & SRD~\cite{srd} \\ 
                        & HPC &  c5n.18xlarge & \thead{2x18C\\Intel Xeon Platinum\\8124M @ 3GHz} & 192 GB & 1.475 USD & 3.88 USD & 100 Gb/s & \thead{Fat Tree~\cite{srd}} & ECMP~\cite{srd} & SRD~\cite{srd} \\ 
    \hline
 \multirow{3}{*}[-3.5ex]{\hspace{2ex}\rotatebox[origin=c]{90}{\textsc{Azure}}} & Normal & F72s\_v2 & \thead{36C\\Intel Xeon Platinum\\8370C/8272CL/8168} & 144 GB & 1.116 USD & 3.045 USD & 30 Gb/s & Fat Tree~\cite{azuredcnet} & ECMP~\cite{azuredcnet} & N.A. \\  
                        
                        & HPC & HC44rs & \thead{2x22C\\Intel Xeon Platinum\\8168 @ 2.70GHz} & 352 GB & 2.218 USD & 3.168 USD & 100 Gb/s & \thead{Non-Blocking\\Fat Tree~\cite{azurevmsizes}}
                        & \thead{Static/\\Adaptive~\cite{azureadaptiverouting}} & InfiniBand~\cite{azurevmsizes}  \\ 
                        
                        & HPC (200 Gb/s) & HB120rs\_v2 & \thead{2x60C\\AMD Epyc\\7V12 @ 2.45 GHz} & 456 GB & 1.8 USD & 3.6 USD & 200 Gb/s & \thead{Non-Blocking\\Fat Tree~\cite{azurevmsizes}}
                        & \thead{Static/\\Adaptive~\cite{azureadaptiverouting}} & InfiniBand~\cite{azurevmsizes}   \\ 
 \hline
 \multirow{2}{*}[-3.5ex]{\hspace{2ex}\rotatebox[origin=c]{90}{\textsc{GCP}}}   & Normal & c2-standard-60 & \thead{2x15C\\Intel Cascade Lake\\@ 3.10GHz} & 240 GB & 1.25 USD & 3.1321 USD & 32 Gb/s & \thead{Jupiter\\(3:1 Blocking\\Fat Tree)~\cite{googlenet}} & ECMP~\cite{googlenet} & \thead{TCP/IP +\\Intel QuickData}~\cite{10.5555/3307441.3307474}  \\  
 
                        & HPC & c2-standard-60 & \thead{2x15C\\Intel Cascade Lake\\@ 3.10GHz} & 240 GB & 2.148 USD & 4.03 USD & 100 Gb/s &  \thead{Jupiter\\(3:1 Blocking\\Fat Tree)~\cite{googlenet}} & ECMP~\cite{googlenet} & \thead{TCP/IP +\\Intel QuickData}~\cite{10.5555/3307441.3307474} \\ 
                        \\
  \hline
  \multirow{2}{*}[-3.5ex]{\rotatebox[origin=c]{90}{\textsc{\thead{Oracle}}}} & \thead{Normal}  & VM.Optimized3.Flex & \thead{18C\\Intel Xeon Gold \\6354 @ 3GHz} & 256 GB & N.A. & 1.188 USD & 40 Gb/s & \thead{Non-Blocking\\Fat Tree~\cite{oracle-clos}} & N.A. & N.A. \\  
  & \thead{HPC (Metal)}       & BM.Optimized3.36 & \thead{2x18C\\Intel Xeon Gold \\6354 @ 3GHz} & 512 GB & N.A. & 2.712 USD & 100 Gb/s & \thead{Non-Blocking\\Fat Tree~\cite{oracle-clos}} & N.A. & RoCEv2~\cite{oracle-roce} \\ 
  \hline
 
 \rotatebox[origin=c]{90}{\textsc{\thead{Daint}}}& \thead{HPC (Metal)}       & - & \thead{2x18C\\Intel Xeon\\E5-2695 v4 @ 2.10GHz} & 64 GB & 1.02 USD~\cite{cscs-pricing} & 1.73 USD~\cite{cscs-pricing} & 82 Gb/s & \thead{Cray Aries\\(Dragonfly)~\cite{alverson2012cray}} & \thead{Per-Packet\\Adaptive~\cite{alverson2012cray}} & FMA~\cite{alverson2012cray} \\
\hline
 \rotatebox[origin=c]{90}{\textsc{\thead{Alps}}}& \thead{HPC (Metal)}       &  - & \thead{2x64C\\AMD EPYC\\7742 @ 2.25GHz} & 256 GB & N.A. & N.A. & 100 Gb/s & \thead{HPE Cray Slingshot\\(Dragonfly)~\cite{slingshot}} & \thead{Per-Packet\\Adaptive~\cite{slingshot}} & RoCEv2~\cite{slingshot} \\ 
\hline
 \rotatebox[origin=c]{90}{\textsc{\thead{DEEP-EST}}}& \thead{HPC (Metal)}       & - & \thead{2x12C\\Intel Xeon Gold\\6146 @ 3.20GHz} & 192 GB & N.A. & N.A. & 100 Gb/s & \thead{Mellanox\\InfiniBand EDR\\(Fat Tree)~\cite{Suarez:905812}} & \thead{Static/\\Adaptive} & Infiniband~\cite{Suarez:905812} \\ 
 \bottomrule
\end{tabular}}
\end{table*}

In this work, we focus on network performance and noise, assessing the impact on performance, scalability, and cost of tightly-coupled HPC communication patterns at scale. Because collecting statistically sound measurements at the scale of thousands of HPC VMs would be too expensive (and on some cloud providers not even feasible), we first perform detailed network performance and noise measurement at small scale. On one side, we analyze this data to spotlight differences in network performance and noise between different cloud and on-premise HPC systems. On the other side, we use this data to calibrate the LogGOPSim simulator~\cite{loggopsim1,loggopsim2}, and to simulate the scalability and cost at a larger scale (up to 16K HPC VMs).

We define the concepts of \textit{latency noise} and \textit{bandwidth noise}, and assess the network performance and its impact on scalability of HPC and normal instances of four major cloud providers and of three on-premise systems (with different network technology). We also assess OS noise (i.e., performance variability introduced by OS processes), and we show how different type of noise impact application performance and cost both at small and large scale, for both latency- and bandwidth-dominated communication patterns.

We describe in Sec.~\ref{sec:hpcinthecloud} the main characteristics of HPC cloud solutions, in Sec.~\ref{sec:perf} we analyze the network performance of both cloud and on-premise HPC systems, with a focus on OS and network noise in Sec.~\ref{sec:noise}. Then, we simulate how noise affects performance at scale in Sec.~\ref{sec:simulations}, and discuss related work in Sec.~\ref{sec:related}. Eventually, Sec.~\ref{sec:conclusions} draws conclusions.

\section{HPC in the Cloud}\label{sec:hpcinthecloud}
In this section we measure and analyze the network performance of HPC systems in the cloud at a small scale, to understand better some peculiarities and limitations of those systems. In this paper, we analyze four of the major cloud providers: Amazon AWS~{\cite{aws}}, Google GCP~{\cite{gcp}}, Microsoft Azure~{\cite{azure}}, and Oracle Cloud~\cite{ocl}. We also analyze three on-premise HPC systems: Piz Daint~\cite{daint} (referred as \textit{Daint} in the following) and Alps~\cite{alps}, both deployed at the Swiss National Supercomputing Centre, and DEEP-EST~\cite{Suarez:905812}, deployed at the J\"ulich Supercomputing Centre. We analyze cloud instances of different types, including HPC instances (with different network bandwidth) and normal compute instances.
We outline the different analyzed systems, instance types, and their main characteristics in Table~\ref{tab:systems}. In the following we analyze in detail the different instance types (Sec.~\ref{sec:systems:inst}), their network features (Sec.~\ref{sec:systems:network}), and their cost (Sec.~\ref{sec:systems:cost}).

\subsection{Instances, CPUs, and OS}\label{sec:systems:inst}
In the following, with the term \textit{HPC instances}, we refer to those instances providing at least 100 Gb/s networking.
For AWS, we evaluate both bare-metal and non bare-metal HPC instances.
Azure and GCP provide only non bare-metal HPC instances, whereas Oracle only provides bare-metal HPC instances. To have a fair comparison, we selected instance types with similar CPUs when possible. We used Intel CPUs on all the cloud instances except for the 200 Gb/s instances of Azure, which only have AMD EPYC CPUs. For normal instances, we selected those that provide a similar network bandwidth and core count. For completeness, we also report the amount of RAM memory on each instance type. 

All four providers guarantee that HPC instances are run on separate physical servers. For the normal instances we selected CPUs with a high core count to have them allocated on two separate servers. This is necessary to ensure that when measuring network performance the two VMs are actually using the network. For the cloud providers we report in the \textit{Instance Type} column the name of the instances we used.

On all cloud providers, we use the virtual machine (VM) images and operating system suggested for the HPC instances. These were: Amazon Linux 2 on AWS~\cite{awshpcimage}, CentOS 7.7 on Azure~\cite{azurehpcimage}, CentOS 7.9 on GCP~\cite{gcphpcimage}, and Oracle Linux 7.9 on Oracle. Daint and Alps run a Cray Linux Environment (CLE) OS based on SUSE Linux Enterprise Server v15.2, and DEEP-EST runs Rocky Linux v8.5.

\subsection{Network}\label{sec:systems:network}
The four cloud providers and the DEEP-EST system deploy a fat tree topology. According to the most recent documentation we found, Azure, Oracle, and DEEP-EST deploy a non-blocking network~\cite{azurevmsizes,deep-est-2}, GCP a 3:1 blocking network~\cite{googlenet}, whereas we did not find any additional detail on network over- or under-provisioning for AWS. Both AWS and GCP use ECMP routing~\cite{ecmp}, Azure employs adaptive routing~\cite{azureadaptiverouting} for HPC instances, and for Oracle we did not find any information on routing.
The routing protocol plays a crucial role in network performance. For example, ECMP is congestion oblivious and might suffer from flow collisions ~\cite{10.1145/2535372.2535375,conga,drill}, increasing the network bandwidth variability (see Sec.~\ref{sec:noise:bw}). Daint and Alps deploy a dragonfly interconnect (Cray Aries~\cite{alverson2012cray} and Slingshot~\cite{slingshot} respectively) with adaptive routing.

Each of the evaluated cloud providers uses a different transport protocol. AWS provides its proprietary RDMA-like protocol called SRD (\textit{Scalable Reliable Datagram})~\cite{srd}, which resembles in some aspects InfiniBand verbs~\cite{ibverbs}. It provides reliable out-of-order delivery of packets and uses a custom congestion control protocol. The AWS Nitro Card~\cite{awsnitro} implements the reliability layer, and the \textit{Elastic Fabric Adapter} (EFA) provides OS-bypass capabilities. To react to congestion, SRD monitors the round trip time (RTT) and forces packets to be routed differently by changing some of the fields used by ECMP to select the path. This approach is probabilistic and might allow avoiding congested paths, but, differently than truly adaptive routing, it does not allow selecting the least congested path nor any specific path.

Azure and DEEP-EST use RDMA through InfiniBand~\cite{azurevmsizes}, Oracle uses \textit{RDMA over Converged Ethernet} (RoCEv2)~\cite{oracle-roce}, whereas
GCP does not use RDMA and relies on TCP/IP. To minimize data movement overheads, GCP uses Intel's QuickData DMA Engines~\cite{quickdata} to offload payload copies of larger packets. Daint uses a proprietary RDMA protocol~\cite{alverson2012cray} (FMA), whereas Alps uses RoCEv2~\cite{slingshot}.

\subsection{Cost}\label{sec:systems:cost}
Table~\ref{tab:systems} shows the per-hour cost charged to the user as of July 18, 2022. For the cloud systems, we report the cost for the East US availability zone. We consider both the cost for a committed 3-years usage with upfront payment and the on-demand cost without any minimum commitment. Please note that 3-years is the maximum commitment allowed on those providers (and that leads to the lowest per-hour cost), whereas when having no commitments we have the highest per-hour cost. For Daint, we report the cost for a minimum usage of $10,000$ compute hours, as well as the on demand cost, both for non-academic partners. Academic partners have discounted rates and this would otherwise lead to an unfair comparison. For Alps and DEEP-EST there is no publicly available information on the per-hour cost. 

On AWS, the main difference between the normal and HPC instances we selected is the support for \textit{Elastic Fabric Adapter} (EFA), which provides the 100 Gb/s networking. Thus, we can estimate the 3-years committed cost of the high-performance networks at around 0.135 USD per hour per VM, and an on-demand cost of 0.82 USD. Similarly, we selected the same instance type for normal and HPC instances on GCP. The only difference is that we enabled the so-called \textit{Tier 1} network on the HPC instance, which provides 100 Gb/s network bandwidth. On GCP, we can thus estimate the cost of the HPC network at around 0.9 USD per hour per VM~\cite{gcp-tier1} (both for the committed and on-demand cost). Unfortunately, Azure and Oracle do not provide the same instance in HPC and non-HPC flavors, and it is thus not possible to isolate the cost of the HPC network from the rest. Also, we observe that whereas the on-demand cost of 100 Gb/s instances is lower than the cost of 200Gb/s instances, this is not true for the 3-years committed usage. Indeed, at the time of the writing, committing for a 3-years usage led to a 30\% discount for 100 Gb/s instances, and to a 50\% discount for the 200Gb/s ones.

\begin{figure*}[htpb]
    \centering
    \includegraphics[width=.9\linewidth]{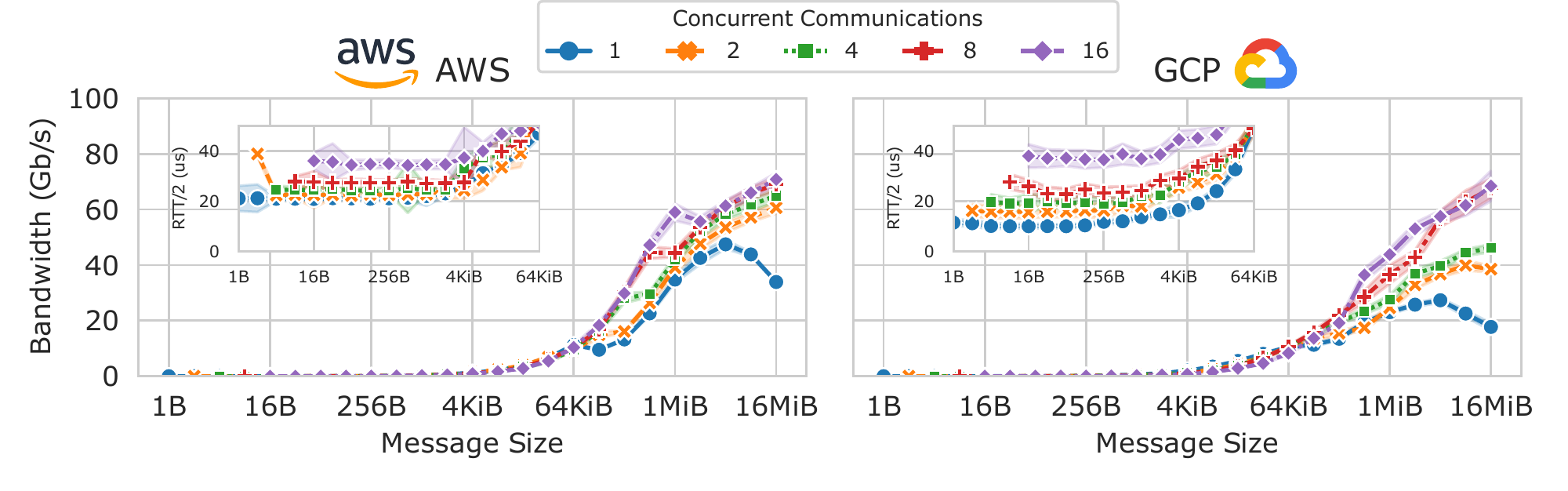}
    \caption{Bandwidth for HPC instances as a function of message size and number of concurrent connections between the two servers. Inner plots show RTT/2 for small messages.}
    \label{fig:conc:bw}
\end{figure*}
\section{Network Performance}\label{sec:perf}
We measure network performance using the \textit{Netgauge} tool~\cite{hoefler-netgauge-hpcc07}, that provides detailed, sample-by-sample measurements (fundamental for estimating network noise in Sec.~\ref{sec:noise}). We used the \textit{Message Passing Interface} (MPI) backend and, on each system, the MPI library recommended by the provider\footnote{On Azure we used HPC-X v2.8.3 on HPC instances and Open MPI v4.1.2 on normal instances. We used Open MPI v4.1.1 on AWS, Intel MPI v2018.4.274 on GCP, Open MPI v4.0.4 on Oracle, Cray MPICH v7.7.18 on Daint, Cray MPICH v8.1.12 on Alps, and Open MPI v4.1.3 on DEEP-EST.}. 

\paragraph{Methods} We created an account on each provider, and used our own funding and/or academic credits, without coordinating with the providers. The clusters have been created and tuned following the guidelines publicly available in the documentation of the cloud providers. After running the benchmarks, we contacted the leads of the cloud business of each of the providers, sharing a draft of the paper with them. They assessed the correctness of our evaluation, and we integrated their feedback in the paper. Only in one case we improved the performance by applying a technique not described in the publicly available documentation, that we describe in the text (see the comment about the \texttt{FI\_EFA\_TX\_MIN\_CREDITS} in Sec.~\ref{sec:bw:saturation}). On all the providers, if not specified otherwise, we allocated the two VMs (or the two servers) on the same rack. The only exception is Oracle, where it is not possible to explicitly control the allocation. We analyze in detail the impact of allocation on performance and noise in Sec.~\ref{sec:noise}. 

\subsection{Bandwidth Saturation}\label{sec:bw:saturation}
All the four analyzed cloud providers claim a 100 Gb/s bandwidth on Intel-based HPC instances. However, this is true only under certain conditions. For example, AWS documents a maximum per-message bandwidth of 25Gb/s~\cite{awsinstancebw}. Even if not explicitly documented, we observed similar limitations on GCP.
One possible reason justifying this behavior is that even if the instance exposes a single 100 Gb/s NIC, it might be equipped with multiple 25Gb/s NICs (or a multi-port NIC). While some providers explicitly documented this for non-HPC instances, the specific configuration is often unclear for HPC ones.



For this reason, we can expect a higher bandwidth when sending a message over multiple connections. To assess if this is the case, we run a ping-pong benchmark between two nodes. We establish multiple concurrent connections between the two nodes, by running multiple processes per node and letting each pair of processes send/receive disjoint parts of the message. For example, a 16MiB pingpong with 16 processes per node runs 16 concurrent ping-pongs between 16 processes on the first node and 16 processes on the second node, each with a 1MiB message.

We report the results of this experiment for AWS and GCP in Figure~\ref{fig:conc:bw}. We report the bandwidth as the message size divided by half the round trip time (RTT/2), and the inner plots show the RTT/2 for small messages. Each point in the plot is the average over 1000 runs, whereas the band around the point represents the standard deviation. We do not report the results for the other systems since they can saturate the bandwidth even with a single connection (we show results in the next section). On AWS we increased the bandwidth by increasing the maximum number of in-flight packets to 1024 (by setting the \texttt{FI\_EFA\_TX\_MIN\_CREDITS} environment variable). 

On both AWS and GCP the bandwidth increases when increasing the number of concurrent communications (up to 80Gb/s with 16 processes per node). Also, when using a single connection, the bandwidth drops for messages larger than 4MiB. This is caused by a more-than-linear increase in last level cache (LLC) misses, that we measured by using the \texttt{perf} tool. For example, on AWS, we observe a 4$\times$ increase in LLC misses when going from 1MiB to 4MiB messages, but a 8$\times$ increase when moving from 4MiB to 16MiB messages. This effect is not present when using more concurrent communications because the message is split among the processes, each transmitting a smaller message. 

We also observe that having more processes per node increases the RTT of small messages, due to additional overhead and contention on the NIC access. For this reason, only large messages should be sent with multiple concurrent connections. Instead of having more processes sending a part of the message each, we could have a single process sending multiple smaller messages. For example, some MPI libraries provide the possibility to \textit{stripe} messages transparently over multiple connections (e.g., by using the \texttt{btl\_tcp\_links} command line flag on Open MPI~\cite{btllinks}). However, we did not observe any performance improvement compared to the single connection case. 

\begin{figure*}[h]
    \centering
    \includegraphics[width=.7\linewidth]{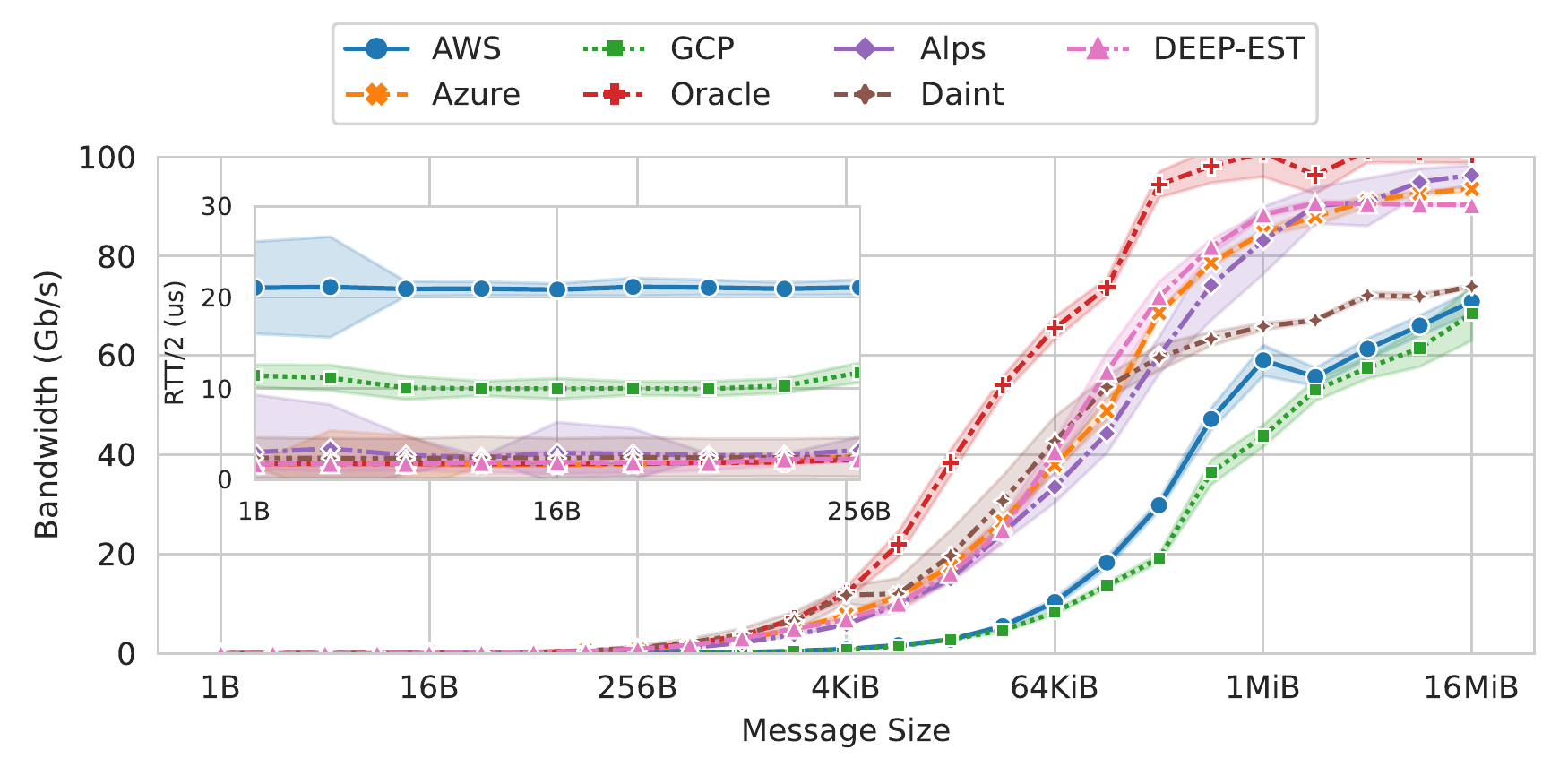}
    \caption{Unidirectional bandwidth on different providers as a function of the message size. Inner plots show RTT/2 for small messages. For AWS and GCP we report the results with optimal number of connections (1 for minimizing RTT/2 for small messages, and 16 for maximizing bandwidth on large messages).}
    \label{fig:latbw}
\end{figure*}

\begin{shaded*}
\textbf{Observation 1:} \textit{On AWS and GCP, the peak bandwidth on a single connection is 50Gb/s and 30Gb/s respectively. A bandwidth of 80Gb/s can only be reached by forcing messages to be concurrently sent/received by/from multiple processes on different connections.}
\end{shaded*}



\subsection{Unidirectional Bandwidth and Latency}\label{sect:uni_bw_lat}
Figure~\ref{fig:latbw} shows the RTT/2 and bandwidth of different providers as a function of the message size. For the cloud providers we selected the 100 Gb/s instances (bare-metal when available). For AWS and GCP, we report the RTT results with a single connection (lowest RTT on small messages), and for bandwidth, the results with 16 concurrent connections (highest bandwidth on large messages). 

Regarding the latency (i.e., the RTT/2 for 1 byte messages), we observe that Azure, Oracle, and the on-premise systems exhibit a 1-2 microseconds latency for HPC instances. On the other hand, both AWS and GCP are characterized by much higher latencies (20 and 10 microseconds respectively). Concerning the bandwidth, Azure, Oracle, Alps, and DEEP-EST achieve a bandwidth higher than 90 Gb/s for 16MiB messages. On Daint, we measured 75 Gb/s peak bandwidth for 16MiB messages (NICs on Daint have an injection bandwidth of 82 Gb/s~\cite{alverson2012cray}). AWS and GCP reach a peak bandwidth of circa 70 Gb/s (using 16 concurrent connections).
Oracle achieves the 90\% of the declared bandwidth with 256KiB messages, Alps, Azure and DEEP-EST with 2MiB messages, Daint with 4MiB messages, and AWS and GCP only achieve the 70\% of the declared bandwidth with 16MiB messages.

\begin{shaded*}
\textbf{Observation 2:} \textit{Azure and Oracle achieve network latency and bandwdith comparable to that of on-premise HPC systems. On the other hand, GCP and AWS achieve 25\% lower bandwidth and 10x higher latency.}
\end{shaded*}

\begin{figure}[htpb]
    \centering
    \includegraphics[width=\linewidth]{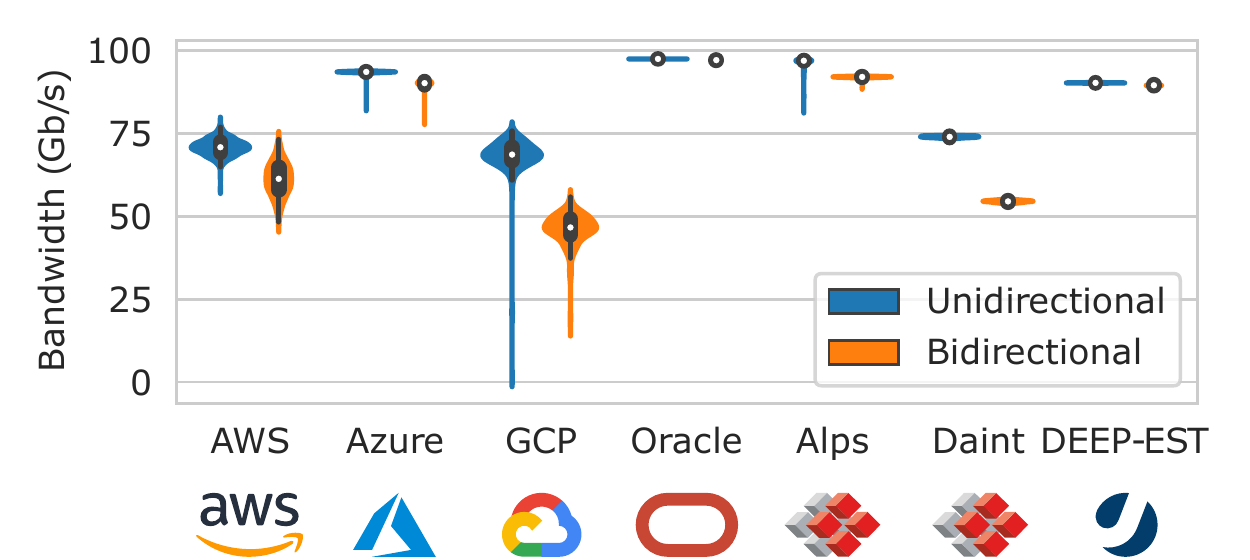}
    \caption{Peak unidirectional and bidirectional bandwidth.}
    \label{fig:uni_vs_bi}
\end{figure}
\subsection{Bidirectional Bandwidth}
To measure the bidirectional bandwidth, we perform two simultaneous ping-pongs between two nodes, with each ping-pong starting from a different node. In Figure~\ref{fig:uni_vs_bi}, we report the results of this experiment, and we compare the peak unidirectional and bidirectional bandwidth with 16MiB messages. For both AWS and GCP, we use 16 concurrent connections. 
In some cases, we observe a peak bidirectional bandwidth lower than the peak unidirectional bandwidth. For example, on Daint this is caused by message requests and responses sharing the same data path, decreasing the peak theoretical bandwidth per direction to 64 Gb/s~\cite{alverson2012cray}.

\begin{figure}[h]
    \centering
    \includegraphics[width=\columnwidth]{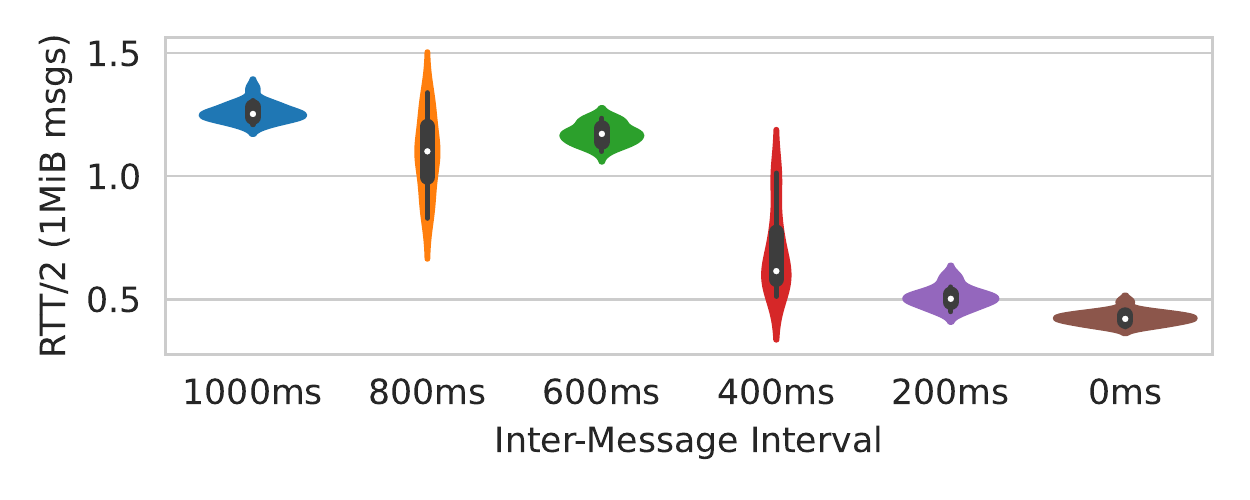}
    \caption{Distribution of RTT/2 (ms) of 1MiB transfers on GCP HPC instances, for different inter-message intervals.}
    \label{fig:hoverboard}
\end{figure}

\begin{figure*}[htpb]
    \centering
    \includegraphics[width=.8\linewidth]{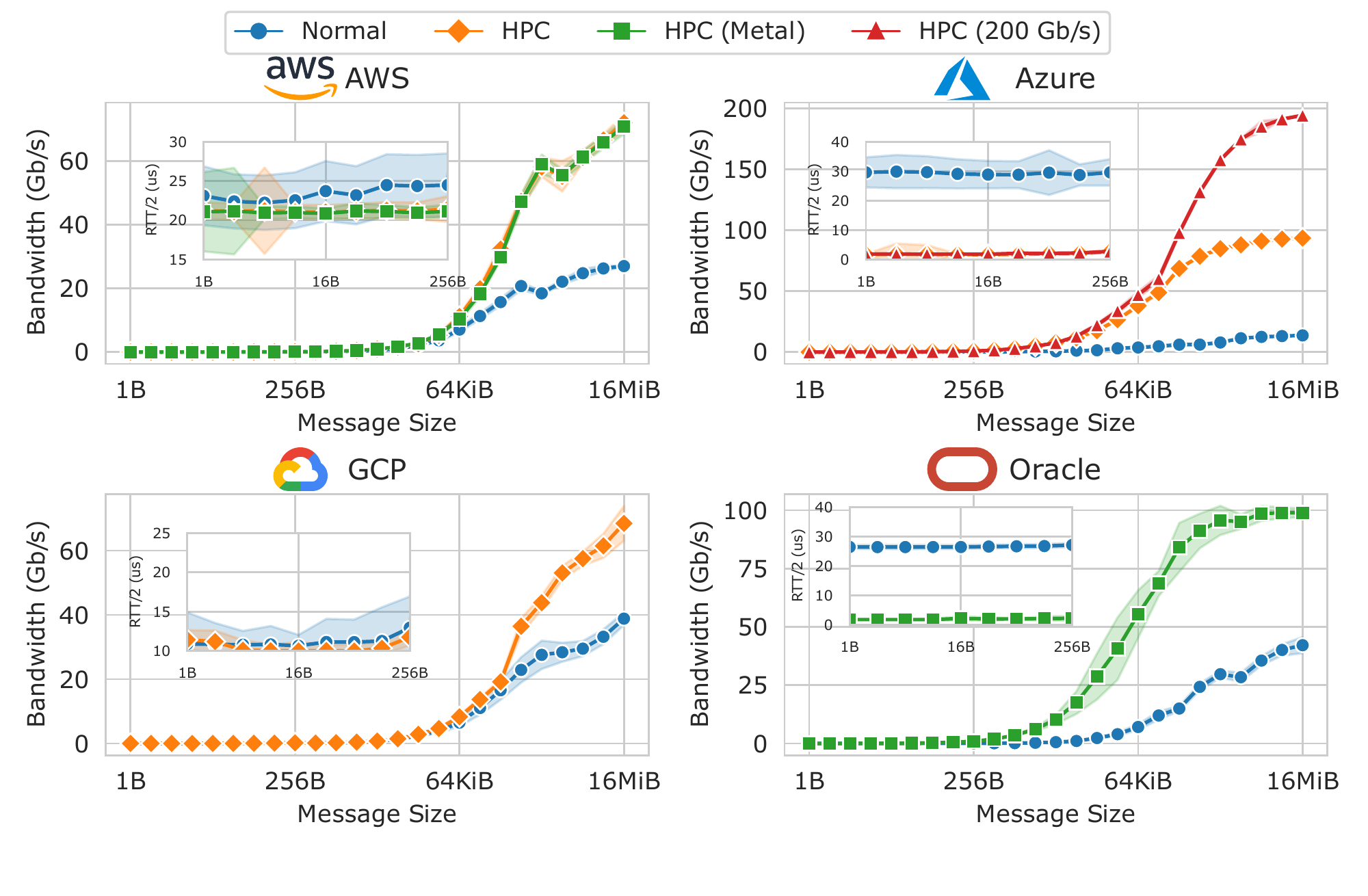}
    \caption{Unidirectional bandwidth for the different instance types described in Table~\ref{tab:systems}, organized by provider. Inner plots show the RTT/2 for small messages. Note that each plot uses a different scale.}
    \label{fig:instances}
\end{figure*}

\subsection{Traffic Burstiness}
We now investigate the impact of traffic burstiness on network performance.
To assess this, we execute a 1MiB ping-pong between two nodes, varying the \textit{inter-message interval}, i.e., the time between two subsequent message transmissions between 0 and 1 second. To exclude any pipelining effect, the benchmark waits for a message to be completely received before sending the next one. We repeat each experiment for 20 iterations, with 10 warm-up iterations. We did not observe any impact of burstiness on the performance, except for GCP, for which we report the results in Figure~\ref{fig:hoverboard}. We show on the X axis the interval between two subsequent messages, and on the Y axis the RTT/2 (milliseconds).

We observe how, when the time interval between subsequent messages is one second, a 1MiB message requires around 1.5 milliseconds to be transferred from the source to the destination. On the other hand, when we decrease the inter-message interval, the RTT starts decreasing, and, eventually, the RTT/2 becomes lower than 0.5 milliseconds. We observed this behavior consistently in multiple runs, in different days, and at different times of the day. 

Our initial assumption was that this could be related to the Andromeda network virtualization stack, used by GCP for forwarding packets over the network~\cite{10.5555/3307441.3307474}. To scale on very large networks, and avoid storing thousands of VM-to-VM forwarding rules on each VM, the Andromeda VM host stack sends all the packets for which it does not have a route to \textit{Hoverboard} gateways. If the Andromeda control plane detects that a flow exceeded some bandwidth usage threshold, it installs direct VM-to-VM forwarding rules in the VM host stack, so that high-bandwidth flows are forwarded directly to the destination VM without the need to traverse the \textit{Hoverboard} gateways to resolve the forwarding rule. Although this works well for bandwidth-intensive flows, bursty flows might not trigger the installation of a forwarding rule in the software stack of the source VM, thus incurring in the extra latency required to traverse a \textit{Hoverboard} gateway. 

However, after a discussion with GCP engineers (that were able to reproduce and confirm the issue) we believe that this is not caused by Hoverboard. Indeed, the same behaviour also happens if the two communicating process run on the same node, thus using shared memory rather than the network for communicating. We also exclude issues with the MPI implementation, since we observed the same behavior when communicating directly using TCP rather than MPI. This is also not caused by CPU power saving features, that were disabled during the tests. We are currently investigating, with the help of GCP engineers, the reasons for this behaviour, that are likely related to virtualization.

\begin{shaded*}
\textbf{Observation 3:} \textit{On GCP, large delays between messages can increase the RTT/2 up to 3x compared to the case where messages are sent back to back.}
\end{shaded*}

\subsection{HPC vs. Normal Compute Instances}
Figure~\ref{fig:instances} reports the unidirectional bandwidth and RTT/2 for the different instance types and cloud providers described in Table~\ref{tab:systems}. For GCP and AWS we report the bandwidth when using 16 concurrent connections. For Azure normal compute instances, we achieved the highest bandwidth (14 Gb/s) with two concurrent connections, and we observed bandwidth degradation when running more than two concurrent connections. We believe that on Azure normal instances concurrent connections are needed because each VM uses multiple NICs with lower bandwidth, as documented by Azure~\cite{azurefsv2}. The Azure 200 Gb/s instances reach the peak bandwidth with 16MiB messages. 
Also, we observe that normal compute instances on Azure are characterized by the lowest bandwidth, whereas AWS normal instances can achieve a 25Gb/s bandwidth, Oracle normal instances achieve a bandwidth of 40Gb/s, and GCP achieve even higher bandwidth than that declared (40Gb/s versus 32Gb/s). 

Regarding the latency, on AWS, HPC instances are characterized by a marginally lower RTT/2 compared to normal instances. On Azure, we observed a 30 microseconds RTT/2 on normal instances, much higher than that observed on the HPC instances (around 1-2 microseconds). The same holds for Oracle, where on normal instances we observed a latency of almost 30 microseconds. On GCP, we observed no difference in the latency between normal and HPC instances.

\begin{shaded*}
\textbf{Observation 4:} 
\textit{On AWS and GCP, HPC instances communicate with the same latency as normal instances, whereas on Azure and Oracle HPC instances communicate with a latency 10-20x lower than normal instances.}
\end{shaded*}

\section{Network and OS Noise}\label{sec:noise}
Application performance can significantly vary across different runs due to effects such as OS and network noise~\cite{10.1145/3295500.3356145,ferreira-os-noise-08, 5645455}. To analyze how these different types of noise can affect the performance of a large scale system, we now describe how they affect the different terms that contribute to the time needed to deliver a message according to the LogGP model~\cite{ALEXANDROV199771}.
We focus on this model because, although simple, it captures the fundamental aspects of network communications, and allows us to implement a solid simulation methodology (discussed in Sec.~\ref{sec:simulations}). LogGP models the time needed for sending a message of $s$ bytes as $T(s) = 2o + L + (s-1)G$, where $o$ is the overhead for sending (and receiving) a message, $L$ is the network latency, and $G$ is the gap between the transmission of two subsequent bytes (i.e., the inverse of the network bandwidth). The model also has an additional parameter $g$ representing the minimum time between the transmission of two subsequent messages. This parameter is not shown in the formula above, because it represents the transmission time of a single message only.

\begin{figure}[h]
    \centering
    \includegraphics[width=\columnwidth]{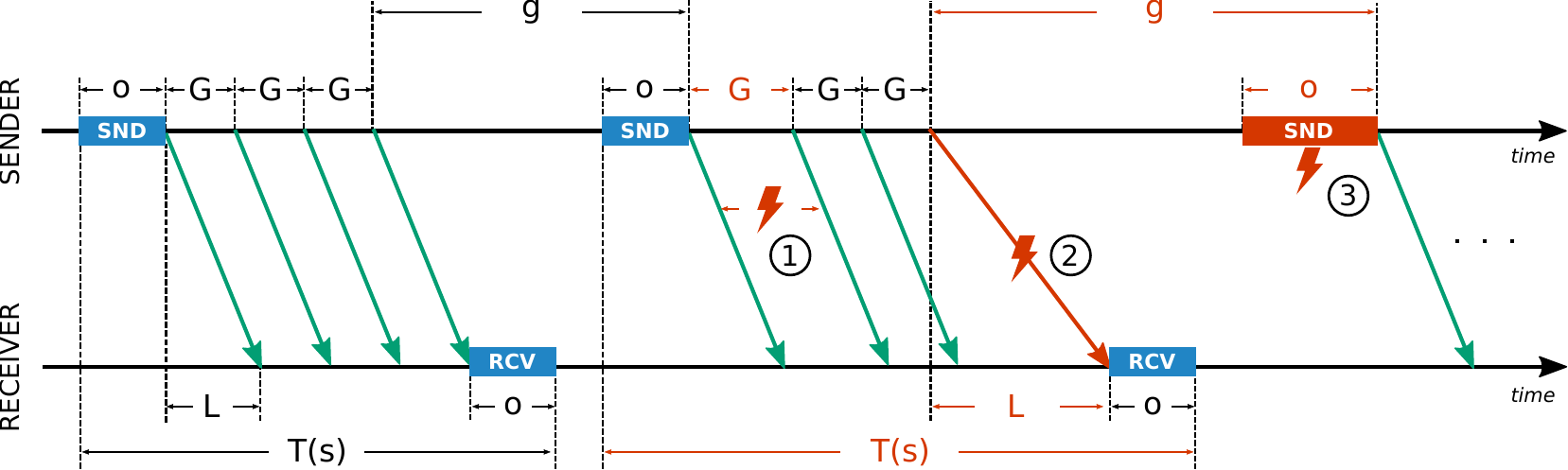}
    \caption{Impact of the different types of noise on the LogGP parameters. \includegraphics[scale=1,trim=0 1 0 0]{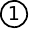} = bandwidth noise, \includegraphics[scale=1,trim=0 1 0 0]{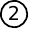} = latency noise, \includegraphics[scale=1,trim=0 1 0 0]{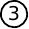} = OS noise.}
    \label{fig:noise}
\end{figure}

To outline the model visually, we show in Figure~\ref{fig:noise} these parameters for a scenario where a node transmits three messages. The first two messages are composed of four bytes, while the third consists of a single byte. Whereas the transmission of the first message is not affected by any type of noise, the transmission of the second message experiences both \textit{bandwidth noise} and \textit{latency noise}. 
Bandwidth noise occurs in the network, and affects the gap per byte $G$, slowing down the transmission of subsequent bytes (or packets) (\includegraphics[scale=1,trim=0 1 0 0]{img/one.pdf}). 
Also latency noise (\includegraphics[scale=1,trim=0 1 0 0]{img/two.pdf}) occurs in the network, and manifests itself as an increase in the latency $L$. It delays the transmission of some bytes (or packets) of the message, and might be caused by transient congestion in the network. The main difference between latency and bandwidth noise is in the duration of the noise events. For example, ECMP routing can cause persistent bandwidth noise due to mapping of multiple network flows on the same paths.
Last, OS noise occurs on the host (\includegraphics[scale=1,trim=0 1 0 0]{img/three.pdf}), can affect both the $o$ and $g$ parameters and, in general, can also cause an increase in the transmission time of a message. The impact of OS noise on the scalability of HPC systems has been largely studied~\cite{ferreira-os-noise-08, 5645455, bhatele13}, but it is still not clear what is its impact on cloud HPC systems, or whether network noise is the major component of cloud systems noise.

\begin{figure}[h]
    \centering
    \includegraphics[width=\columnwidth]{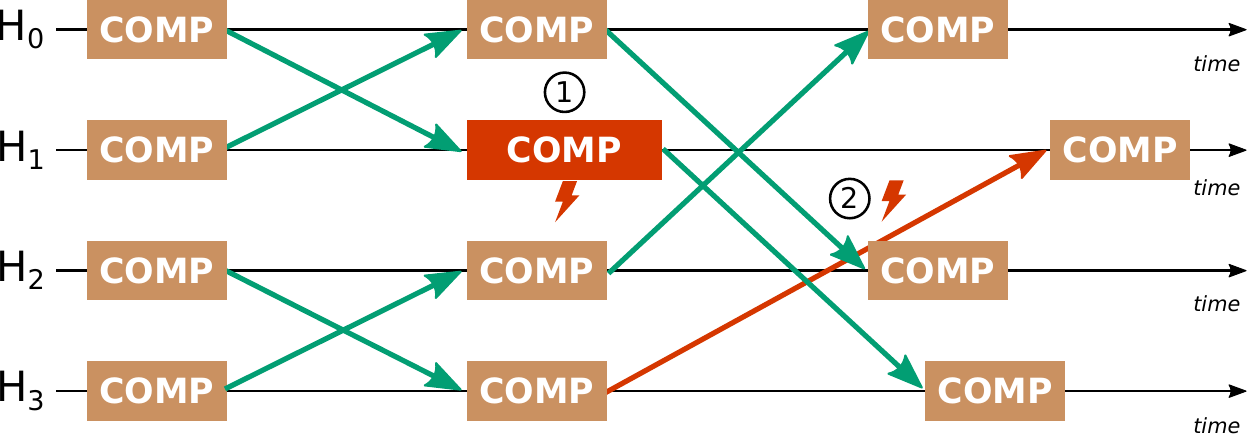}
    \caption{Noise affecting a tightly coupled butterfly-structured collective operation. \includegraphics[scale=1,trim=0 1 0 0]{img/one.pdf} = OS noise, \includegraphics[scale=1,trim=0 1 0 0]{img/two.pdf} = latency noise}
    \label{fig:noise-collectives}
\end{figure}

Moreover, the larger the application scale, the higher the probability that some data transmissions experience either OS or network noise. This can be an issue in tightly coupled applications, where it is enough to delay one single process to slow down the entire application, thus limiting its scalability. Moreover, multiple noise events can either overlap, or sum up and amplify. 
We exemplify this in Figure~\ref{fig:noise-collectives}, showing a few steps of a collective operation performing a butterfly communication pattern. In this example, one process is affected by OS noise, thus delaying the transmission of its data (\includegraphics[scale=1,trim=0 1 0 0]{img/one.pdf}), while another process is affected by latency noise (\includegraphics[scale=1,trim=0 1 0 0]{img/two.pdf}). Because these two noise events overlap, the application is only delayed by the maximum between the two. In general, non-overlapping noise events can accumulate and significantly increase the application execution time. In the remaining part of this section, we analyze the noise events we observed on the systems of Table~\ref{tab:systems}. Then, in Sec.~\ref{sec:simulations} we simulate how these events affect applications at scale.
\begin{figure*}[htpb]
    \centering
    \includegraphics[width=\linewidth]{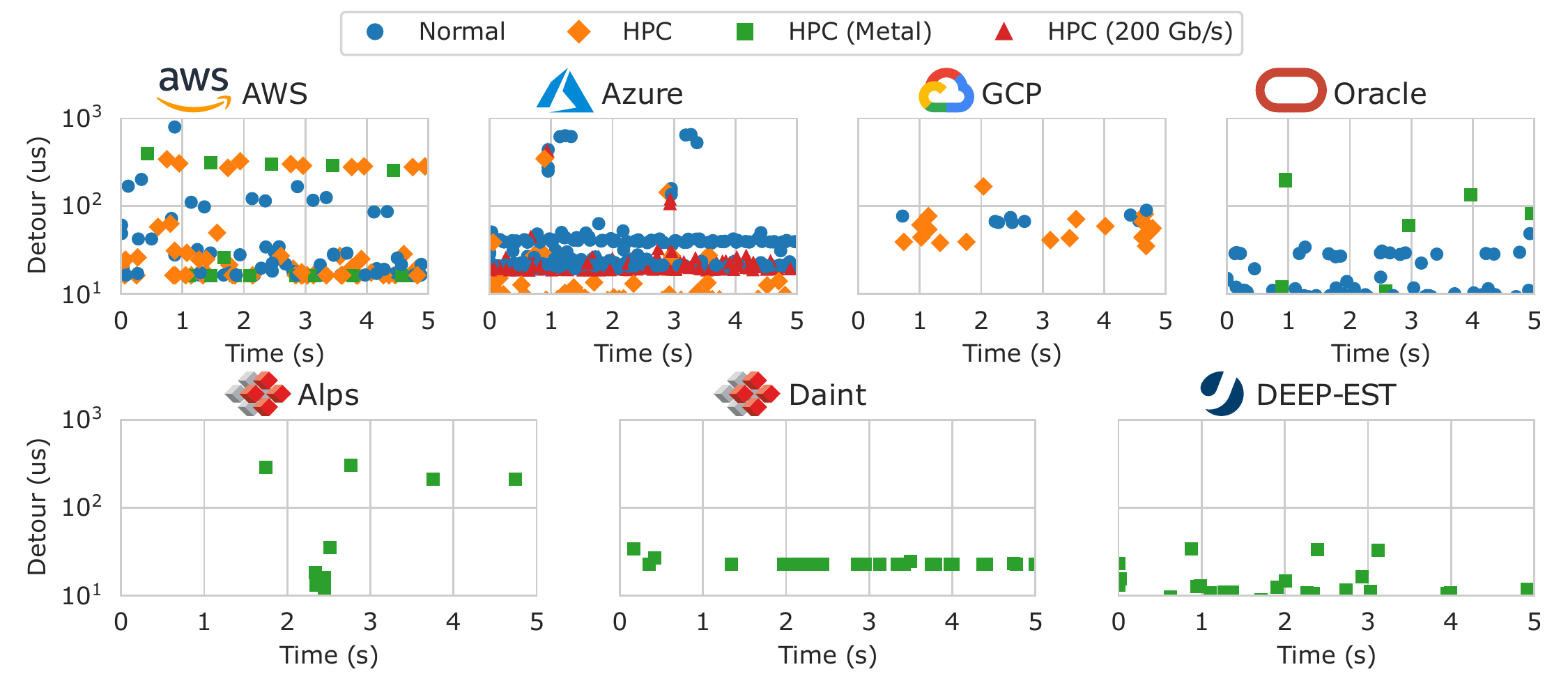}
    \caption{OS noise for the different instance types described in Table~\ref{tab:systems} for the different providers.}
    \label{fig:noise:instances}
\end{figure*}

\subsection{OS Noise}\label{sec:noise:os}
We measured OS noise by running the \textit{selfish detour} benchmark provided by Netgauge, using the same parameters suggested by the authors~\cite{5645455}. The benchmark measures perturbation introduced by the OS, and has been successfully used in the past to assess OS noise of some on-premise HPC systems~\cite{5645455}. It runs a tight loop and records all the iterations larger than $9\cdot t_{min}$, where $t_{min}$ is the minimum measured time required to complete one iteration. The benchmark runs until a predefined number of iterations are recorded. Figure~\ref{fig:noise:instances} shows detour durations over time for a 5 seconds window, for all considered providers. To improve the readability of the plots, we only report the largest samples (top 1\%). It is worth remarking that those largest samples are exactly those representing the noise and limiting applications scalability~\cite{5645455}.

We observe that on-premise systems have the lowest OS noise, as well as HPC bare-metal instances on AWS and Oracle. On the other hand, on AWS and Azure, HPC and normal instances experience noise with high intensity and high frequency, whereas GCP experience noise with medium frequency and low intensity. On Azure, normal and 200 Gb/s instances experience noise with higher frequency than the 100 Gb/s HPC instances.

\begin{shaded*}
\textbf{Observation 5:} \textit{When available, bare-metal HPC instances are characterized by a lower OS noise than non bare-metal instances, and are comparable to on-premise HPC systems. Normal instances are characterized by OS noise with high frequency.}
\end{shaded*}

\subsection{Latency Noise}\label{sec:noise:lat}
We now evaluate the impact of latency noise by running a 1-byte ping-pong for one hour. 
Generally, the performance of a multi-tenant computing system might change depending on the time of the day (due to the different system utilization).
We collected data for 24 consecutive hours on all the analyzed systems (not shown here due to space constraints), and we did not observe significant intra-day variability. For consistency, we report here the data collected at 5 PM in the local time of the cluster. Performance might also change depending on the distance between the two nodes in the network. For this reason, we analyze the noise for different node distances. On AWS and GCP it is possible to specify whether the VMs in a cluster must share the same rack (and thus the same network switch) or not. We did the same on the three on-premise systems for consistency and fairness. On Azure, unfortunately, to use the high-bandwidth network, HPC VMs must always share the same rack. On Oracle it is not possible to specify or check the allocation. However, to simplify the exposition of the results, we assume that the two VMs share the same rack.

\begin{figure*}[htpb]
    \centering
    \includegraphics[width=\linewidth]{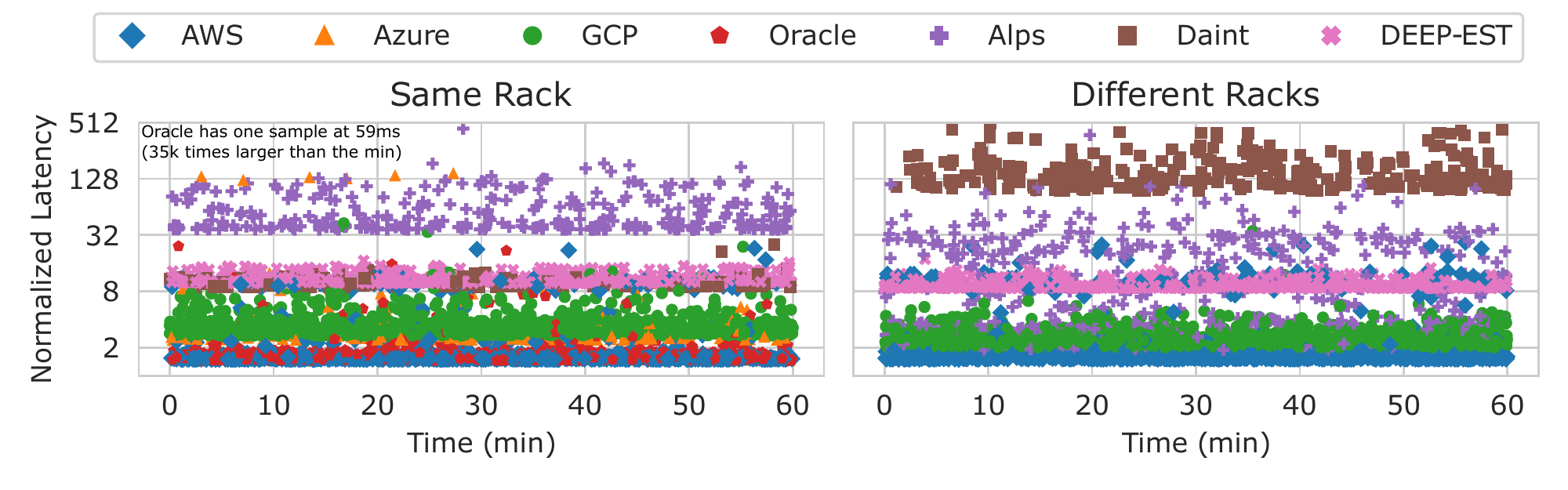}
    \caption{Latency noise for different node distances for 100 Gb/s HPC instances. Base latency is reported in Table~\ref{table:data}.}
    \label{fig:noise:lat:timealloc}
\end{figure*}

Figure~\ref{fig:noise:lat:timealloc} reports the results of our latency noise assessment on the different providers for different node distances for 100 Gb/s HPC instances (bare-metal when applicable). We show the latency normalized with respect to the minimum latency (for completeness, we report in Table~\ref{table:data} the minimum and average latency). To improve the readability of the plot, we only report the largest 0.1\% measurements.

\begin{table*}[htpb]
\centering
\caption{Minimum and average latency and bandwidth for the different providers, instance types, and node distances.}
\label{table:data}
\resizebox{\linewidth}{!}{
\begin{tabular}{p{.5cm} l | c c | c c c | c c | c c | c | c | c} 
 & & \multicolumn{2}{c|}{AWS} & \multicolumn{3}{c|}{Azure} & \multicolumn{2}{c|}{GCP} & \multicolumn{2}{c|}{Oracle} & Alps & Daint & DEEP-EST \\ 
 & & Normal & \thead{HPC\\(Metal)} & Normal & HPC & \thead{HPC\\(200 Gb/s)} & Normal & HPC & Normal & \thead{HPC\\(Metal)} & \thead{HPC\\(Metal)} & \thead{HPC\\(Metal)} & \thead{HPC\\(Metal)} \\
 \midrule \hline
 \multirow{4}{*}{\rotatebox[origin=c]{90}{\textsc{\thead{Same\\Rack\\}}}} & Min. Lat. (us) & 19.28 & 16.79 & 26.11 & 1.50 & 1.70 & 9.42 & 8.46 & 24.68 & 1.66 & 2.13 & 1.19 & 1.19 \\ 
 & Mean Lat. (us) & 23.11 & 18.97 & 29.59 & 1.65 & 1.84 & 10.64 & 9.98 & 26.43 & 1.72 & 3.01 & 2.39 & 1.70 \\
 & Max. Band. (Gb/s) & 27.32 & 78.74 & 7.42 & 93.92 & 194.48 & 45.45 & 75.75 & 11.20 & 97.53 & 97.14 & 74.37 & 90.46 \\ 
 & Mean Band. (Gb/s) & 27.02 & 70.84 & 7.28 & 93.51 & 194.25 & 38.82 & 68.45 & 8.39 & 97.50 & 96.32 & 73.93 & 90.27 \\ \hline
 \multirow{4}{*}{\rotatebox[origin=c]{90}{\textsc{\thead{Different\\Racks\\}}}} & Min. Lat. (us) &  22.46 & 17.20 & N.A. & N.A. & N.A. & 12.39 & 14.90 & N.A. & N.A. & 2.66 & 1.19 & 1.41 \\ 
 & Mean Lat. (us) & 27.57 & 19.26 & N.A. & N.A. & N.A. & 15.02 & 16.66 & N.A. & N.A. & 2.90 & 3.33 & 1.93 \\
 & Max. Band. (Gb/s) & 30.52 & 77.72 & N.A. & N.A. & N.A. & 34.84 & 70.11 & N.A. & N.A. & 96.15 & 75.24 & 90.49 \\ 
     & Mean Band. (Gb/s) & 30.14 & 67.02 & N.A. & N.A. & N.A. & 30.67 & 65.71 & N.A. & N.A. & 96.00 & 74.59 & 90.26 \\
 \bottomrule
\end{tabular}
}
\end{table*}


We observe that, when the two nodes are on the same rack, Alps experiences latency noise with the highest intensity, with samples experiencing a latency more than 128 times higher than the minimum. Also, on Oracle we observed one measurement equal to 59 milliseconds, 35000 times larger than the minimum latency. The other systems instead experience noise with lower intensity. When the two nodes are on different racks, the noise significantly increases on Daint. One of the reasons is adaptive routing, that can sometimes unnecessarily select longer paths~\cite{10.1145/3295500.3356196}. 

On-premise HPC systems are characterized by latency noise with higher intensity, due to the generally lower latency (cf. Sec.~\ref{sec:perf}). As a consequence, fluctuations in the latency have a larger relative impact (but a smaller absolute impact). It is worth remarking that, regardless of the absolute impact of noise on the latency, as we will show in Sec.~\ref{sec:simulations}, noise at scale can negatively impact the performance, even on systems characterized by a lower base latency.

\begin{figure*}[htpb]
    \centering
    \includegraphics[width=\linewidth]{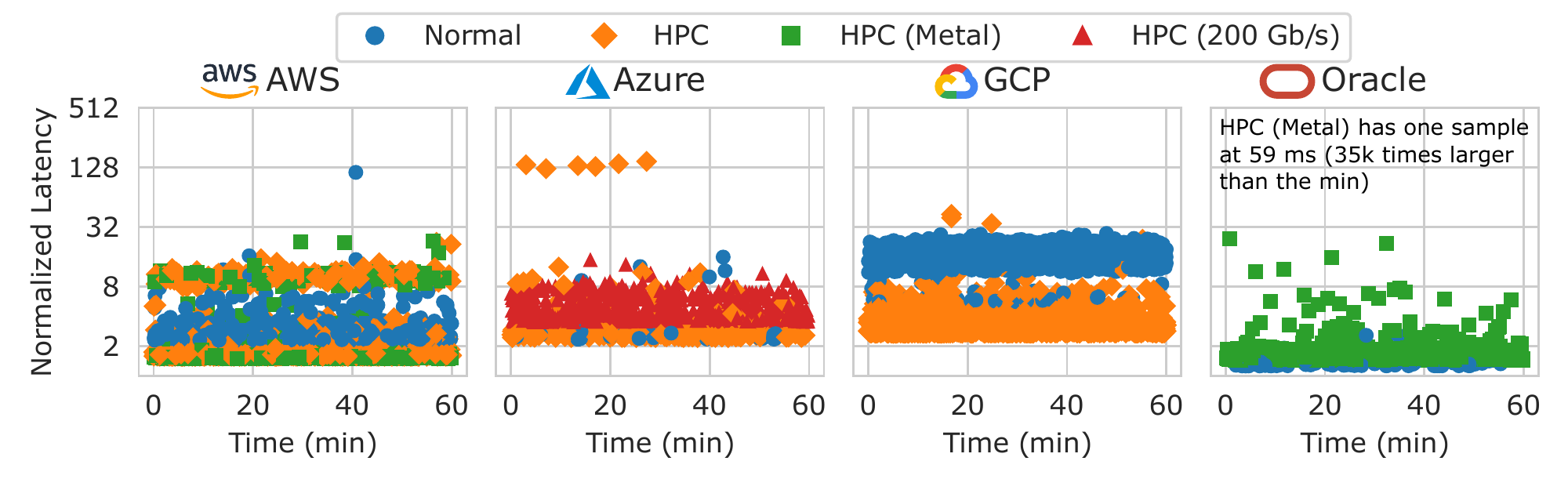}
    \caption{Latency noise for different instance types. Base latency is reported in Table~\ref{table:data}.}
    \label{fig:noise:lat:instancetypes}
\end{figure*}

Figure~\ref{fig:noise:lat:instancetypes} shows the latency noise for different instance types for the four cloud providers (with instances running in the same rack). We observe that on GCP normal instances are characterized by higher noise compared to HPC instances, whereas, on Azure, HPC instances experience a few high-intensity noise events. We did not observe significant differences between the different instance types on AWS, whereas on Oracle we observe a higher noise on HPC instances compared to normal instances. 

\begin{figure*}[htpb]
    \centering
    \includegraphics[width=\linewidth]{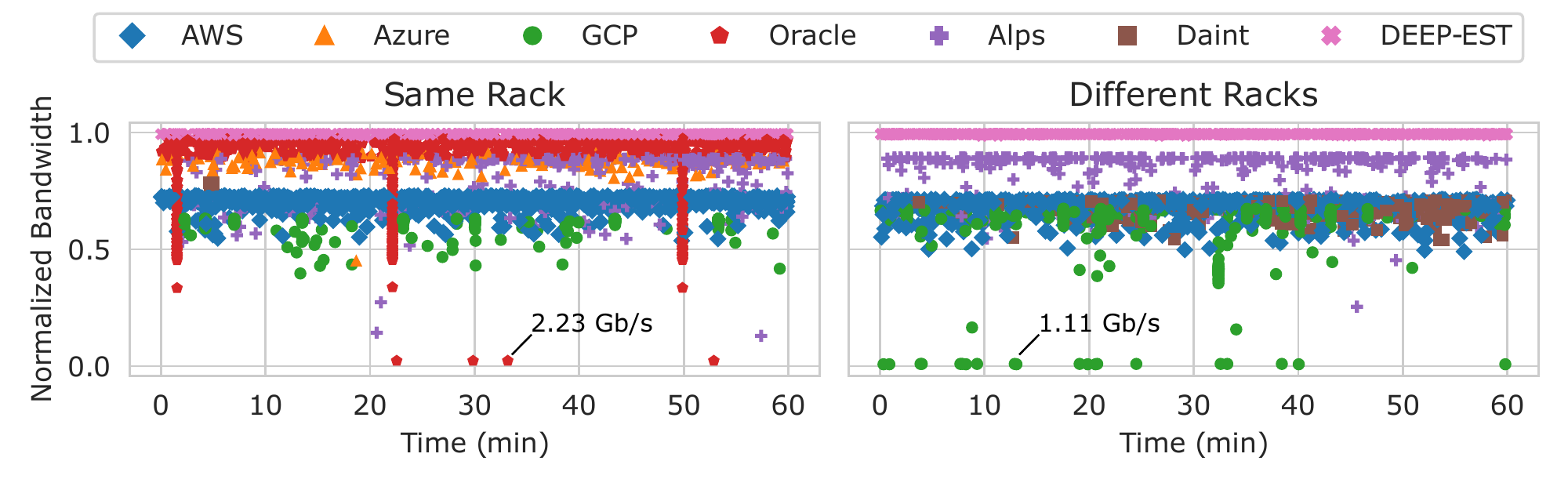}
    \caption{Bandwidth noise for different node distances for 100 Gb/s HPC instances. Base bandwidth is reported in Table~\ref{table:data}.}
    \label{fig:noise:bw:timealloc}
\end{figure*}

\begin{figure*}[htpb]
    \centering
    \includegraphics[width=\linewidth]{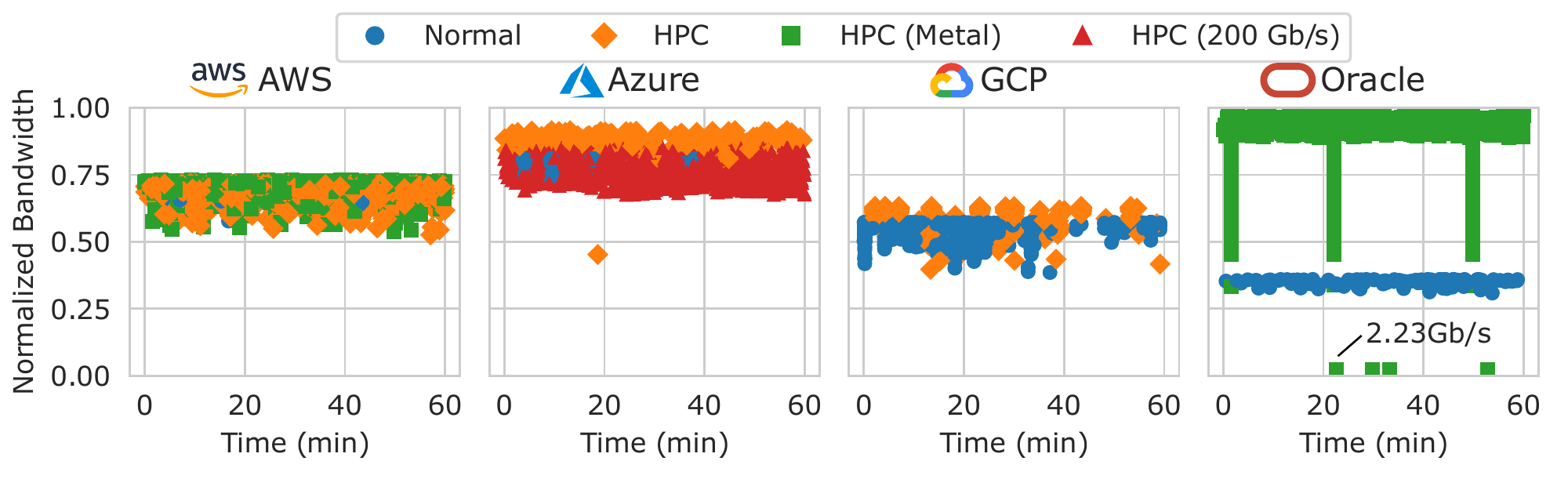}
    \caption{Bandwidth noise for different instance types. Base bandwidth is reported in Table~\ref{table:data}.}
    \label{fig:noise:bw:instancetypes}
\end{figure*}

\begin{figure*}[htpb]
    \centering
    \includegraphics[width=\linewidth]{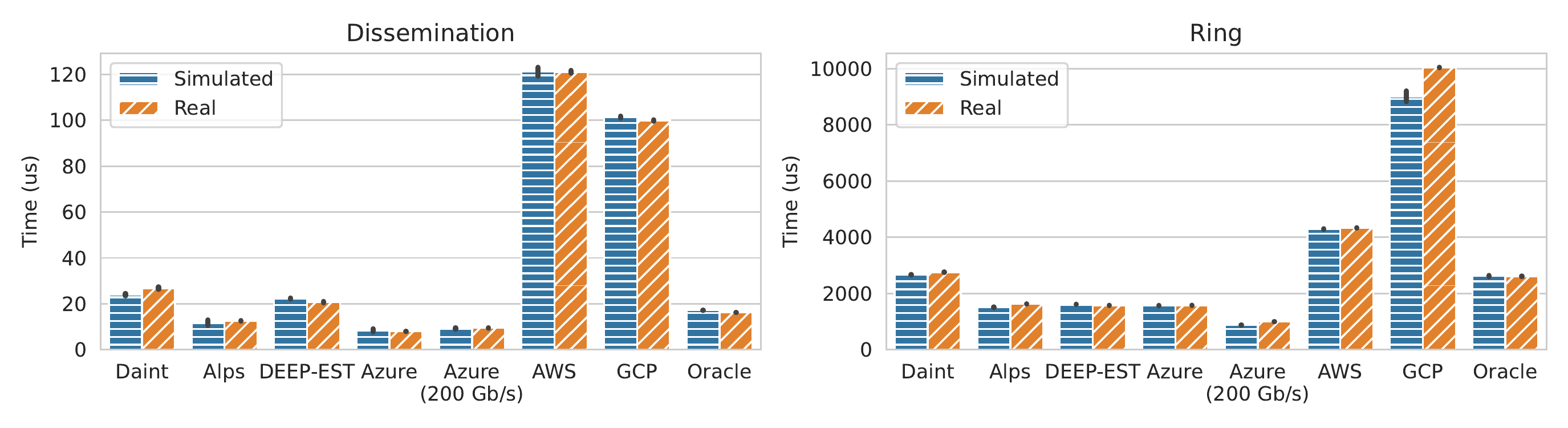}
    \caption{Comparison between measured and simulated times (on HPC instances) for 16B dissemination and 16MiB ring collectives on 16 nodes. Vertical lines at the top of the boxes represent the 95\% confidence interval.}
    \label{fig:simulation:validation}
\end{figure*}

\begin{shaded*}
\textbf{Observation 6:} Latency noise affects both cloud and on-premise HPC systems, and can increase the latency by more than 100x (in a single case up to 35000x). Except that for GCP, HPC instances are not characterized by a lower latency noise than normal instances.
\end{shaded*}

\subsection{Bandwidth Noise}\label{sec:noise:bw}
To measure bandwidth noise, we run large message ping-pongs between two nodes for one hour, and record each sample. For each provider, we select the size of the message that saturates the bandwidth, as well as the optimal number of connections. Even in this case we did not observe significant intra-day variability.  

We report the bandwidth noise for HPC instances (bare-metal when applicable) for different node distances in Figure~\ref{fig:noise:bw:timealloc}. We normalize the bandwidth with respect to the maximum and, to improve the readability of the plot, we only plot the bottom 0.1\% samples (i.e, those with the lowest bandwidth). For completeness, we report in Table~\ref{table:data} the maximum and average bandwidth. We observe that, when the two nodes are on the same rack, AWS, GCP, and Oracle are the most affected by bandwidth noise. On Oracle, bandwidth noise is clustered in time, causing persistent drops that last for up to 0.6 seconds.

Daint and DEEP-EST are almost unaffected by bandwidth noise, whereas Azure and Alps experiences some bandwidth drops. When the two nodes are on different racks, almost all the systems (except for DEEP-EST and Azure) experience an increase in bandwidth noise, with GCP having a few samples experiencing a severe bandwidth drop.

Figure~\ref{fig:noise:bw:instancetypes} shows the bandwidth noise for the different instance types, when the two nodes are on the same rack. We do not observe significant differences between the different instance types on AWS and GCP. On Azure, we observe a slightly higher bandwidth noise on 200 Gb/s instances, for the same reason discussed for the latency noise (small congestion events have a larger relative effect at higher bandwidth). On Oracle, normal instances are severely affected by bandwidth noise, with many samples characterized by a bandwidth equal to 40\% of the maximum achievable bandwidth.

\begin{shaded*}
\textbf{Observation 7:} Cloud systems are more affected by bandwidth noise than on-premise systems. When nodes are on two different racks, the impact of bandwidth noise increases on GCP and Daint. On GCP and Oracle, normal instances are more affected by bandwidth noise.
\end{shaded*}

\section{Large-Scale Simulations}\label{sec:simulations}
To assess the impact of noise at scale, we simulate the performance of collective operations with the LogGOPSim simulator~\cite{loggopsim1,loggopsim2}.  This allows us to analyze the impact of different types of noise in a controlled environment and to estimate the scalability of distributed applications at scale.

\subsection{Simulation Setup and Validation}\label{sec:simulations:validation}
LogGOPSim uses the LogGOPS network model to simulate the execution of parallel algorithms and entire applications. The simulator takes as input the LogGOPS parameters and a program to simulate on an arbitrary number of nodes. The program can either be specified  through the \textit{Group Operation Assembly Language} (GOAL~\cite{hoefler-goal}) or through MPI traces. GOAL specifications are composed of a series of send and receive operations for each process, plus synthetic computational operations, and dependencies among these operations. 
The simulator can also take as an input an OS noise trace, and we extended it to use latency and bandwidth noise traces. Every time a message is sent, instead of simulating a fixed term for the latency or the bandwidth, a value is drawn from the latency and bandwidth noise distributions measured on the real system. Those distributions are built according to the measurement taken with Netgauge (see Sec.~\ref{sec:noise}).

Before simulating the impact of noise at a large scale, we validate the simulator on different collective algorithms. We implemented a program that generates the MPI code of a specific distributed algorithm starting from its GOAL specification. We then execute the generated code on 16 nodes (due to quota limitations, this is the largest allocation that we could get on all the cloud providers at the time of writing.) on the HPC instances. We show in Figure~\ref{fig:simulation:validation} the time measured on the real systems and the time simulated by LogGOPSim starting from the corresponding GOAL trace.

We consider two different widely-used collective operations: a dissemination algorithm (like those used in barriers and small allreduce operations) and a ring-based collective (like the one used in reduce-scatter, allgather, and allreduce operations on large data). Dissemination is executed with 16B messages, and ring with 16MiB messages. We observe how the simulated time closely match the measured time, with a relative error lower than 10\% on both collectives. Similar results have been observed on 4, 8, and 32 nodes, and shown in Appendix~\ref{sec:appendix:validation}. We also run the validation up to 128 nodes on Daint (not shown in the plot), observing a relative error below 5\%.

\subsection{Analysis of Noise on Dissemination Collectives}\label{sec:sim:noise:diss}
\begin{figure*}[htpb]
    \centering
    \includegraphics[width=\linewidth]{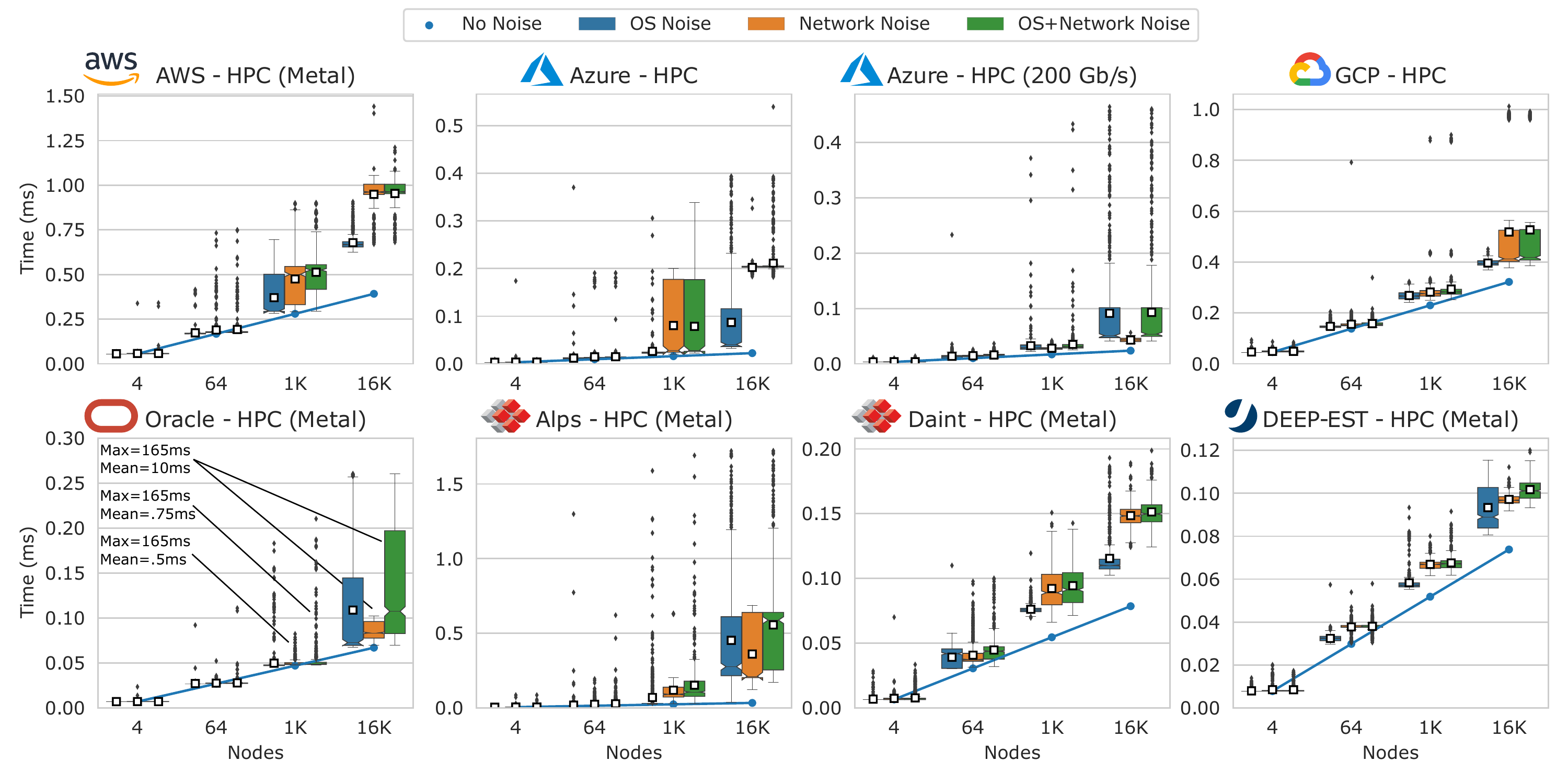}
    \caption{Simulation of the scalability of a 16B dissemination algorithm, with and without OS and network noise. Y-axes have different scales.}
    \label{fig:simulation:dissemination}
\end{figure*}

To analyze the impact of different types of noise on the scalability of parallel applications, we simulated the performance of a 16B dissemination collective on the different systems, and we report the results in Figure~\ref{fig:simulation:dissemination}. For the cloud providers, we only consider HPC instances. We report the results without any noise, with only OS noise or only network noise, and with both noise types. For AWS, we do not report the results for HPC non bare-metal instances, because we did not observe any difference from HPC bare-metal. 

We use different scales for the Y-axes in each subplot for readability purposes. We repeated each experiment 1000 times, and we report the entire distribution of the samples as boxplots. In each boxplot, the middle line represents the median, the notch around the median is the 95\% confidence interval, and the small white square indicates the mean. The upper and lower limits of the box indicate the first quartile (Q1) and the third quartile (Q3). Being $IQR = Q3-Q1$ the interquartile range, the lower and upper whiskers indicate the smallest sample $> Q1-1.5 \cdot IQR$, and the largest sample $< Q3 + 1.5 \cdot IQR$, respectively. Samples outside the whiskers are outliers and reported as black diamonds.

We first focus on the performance in the ideal case where noise does not affect the systems.  We observe that, even without noise, both AWS and GCP are characterized by a lower performance than Azure, Oracle, and the on-premise systems, up to an order of magnitude lower. This is due to the latency on AWS and GCP being one order of magnitude larger than that of the other systems (see Table~\ref{table:data}). 
Starting from 1K nodes AWS performance is severely affected by noise. As highlighted in Sec.~\ref{sec:noise:os}, AWS systems are those experiencing OS noise with the highest intensity, and this significantly increases the runtime of the collective operation, up to 2x on 16K nodes. Even without any OS noise, network noise would still increase the runtime by 50\%. We did not observe amplification effects when adding both types of noise.

Something similar occurs on Azure HPC instances, where network noise increases the average runtime by more than a factor of 4 on 16K nodes, with a few outliers increasing the runtime by 10 times compared to a noiseless execution. 
On the other hand, 200Gb/s instances are mostly affected by OS noise since, as shown in Sec.~\ref{sec:noise:os}, latency noise has lower intensity compared to 100Gb/s instances.
GCP is characterized by network noise with lower intensity, reflecting on a lower relative increase in the runtime.

On Oracle, we observe a small increase in the median runtime. However, a few outliers (not shown in the plot for the sake of readability) experienced a runtime of 165ms on 16K nodes, increasing the average runtime to 10ms. These outliers are caused by some rare noise events with very high intensity (as discussed in Sec.~\ref{sec:noise:lat}). OS and network noise severely affect on-premise systems as well, with Daint experiencing a 2x increase in the runtime on 16K nodes due to OS noise, and up to almost 4x due to network noise.

\begin{shaded*}
\textbf{Observation 8:} OS and network noise can increase the median runtime up to 2x on 1K nodes, and up to 10x on 16K nodes, both on on-premise and cloud systems.
\end{shaded*}

\subsection{Analysis of Noise on Ring Collectives}
We also analyze the impact of noise on bandwidth-dominated ring collectives. This type of collectives are widely used in the training of deep learning models to average gradients~\cite{10.1145/3320060}. In a ring collective each node iteratively receives a chunk of data from its left neighbor, and sends another chunk of the same size to the right neighbour. An allreduce performed as a ring collective is bandwidth optimal, and particularly effective on a smaller node count and for large data. However, due to the long chain of dependencies (each node has to wait to receive data from the previous node before sending), it is enough to slow down a single message transmission to delay the entire operation. 

\begin{figure*}[htpb]
    \centering
    \includegraphics[width=\linewidth]{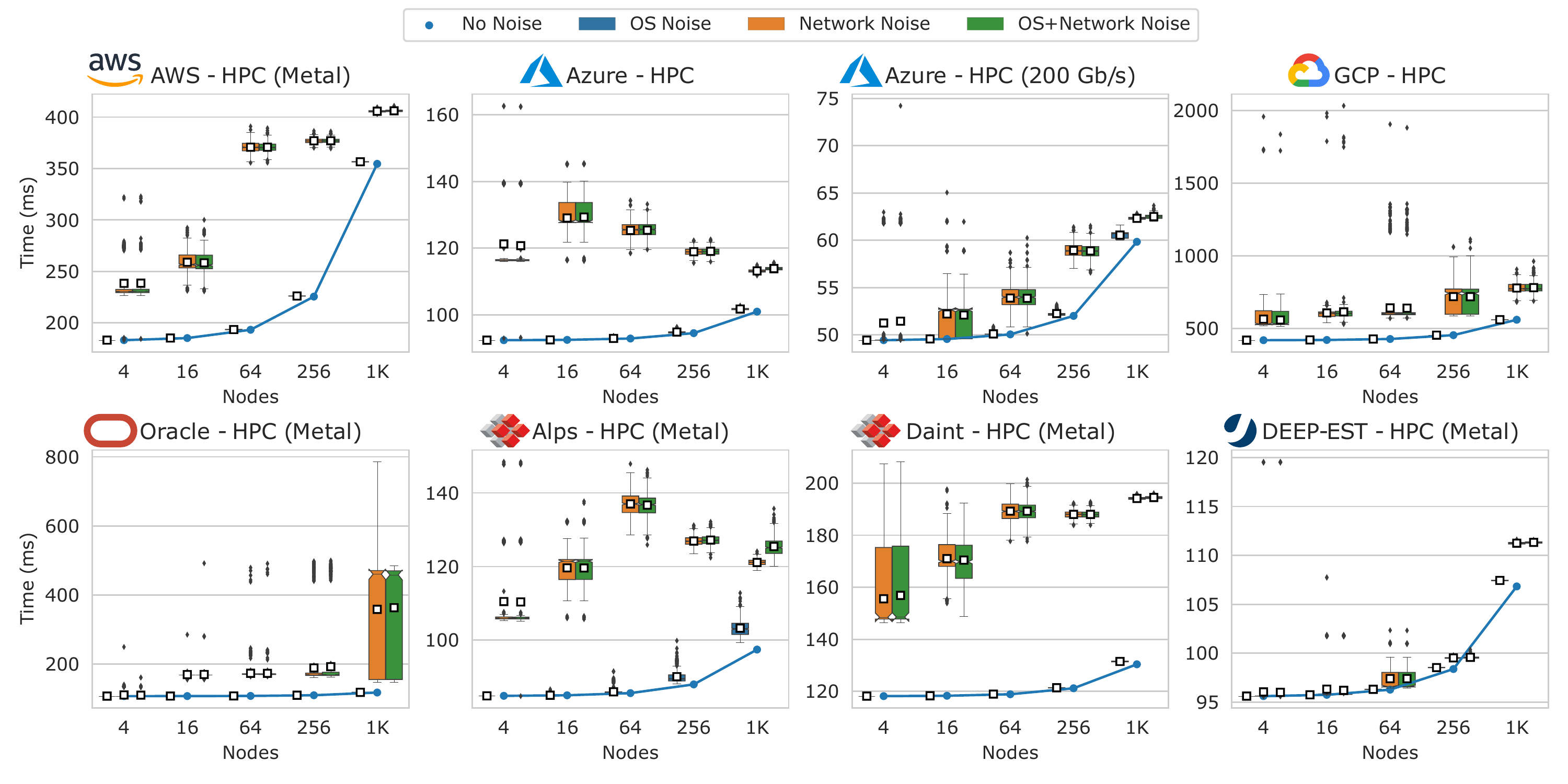}
    \caption{Simulation of the scalability of a 512MiB ring allreduce collective on HPC instances.}
    \label{fig:simulation:ring}
\end{figure*}

For this reason, we show in Figure~\ref{fig:simulation:ring} the results of the simulation of a 512MiB ring allreduce collective. First, for all the systems we observe no impact of OS noise on the runtime. Also, we observe how network (bandwidth) noise impacts the performance also on when running the application on 4 nodes only. 
On almost all the providers we observe network noise increasing the runtime by almost 50\% on 4 nodes. On GCP, a few outliers increase the runtime up to 5 times on 16 nodes, whereas on Oracle the noise increases the average runtime by 4x on 1024 nodes.

\begin{shaded*}
\textbf{Observation 9:} Bandwidth noise can severely affect both on-premise and cloud systems, increasing the runtime by 50\% even when running at small scale (4 nodes).
\end{shaded*}

\subsection{Analysis of Noise Impact on Cost}

Last, we analyze the cost of running HPC workloads in the cloud, by also factoring in the costs due to the different CPUs used by the different providers. We simulate a strawman application running a $128\times128$ double-precision matrix-matrix multiplication followed by a dissemination collective. The message size in the dissemination phase is 128 KiB, equal to the size of the matrix ($128\times128\times 8$ bytes). This communication pattern represents applications that perform some computation followed by reductions (e.g., deep learning training workloads). We measured the time required to perform matrix multiplications on the different CPUs deployed in the analyzed systems by running a benchmark~\cite{gemmbench} using MKL on Intel processors, and BLIS on AMD processors. 

It is worth remarking that the purpose of this benchmark is not to extensively evaluate CPUs performance (that is outside the scope of this paper), but rather to factor in our cost estimation the difference in costs due to different CPU technology. Also, collecting traces of real applications and then simulating them would complicate the interpretability of the results. For example, differences between systems due to differences in OS, libraries compilers, and tools would be amplified on complex applications. By using a mockup application, we can instead keep these differences at a minimum and spotlight differences due to the network rather than on other factors.

\begin{figure*}[htpb]
    \centering
    \includegraphics[width=\linewidth]{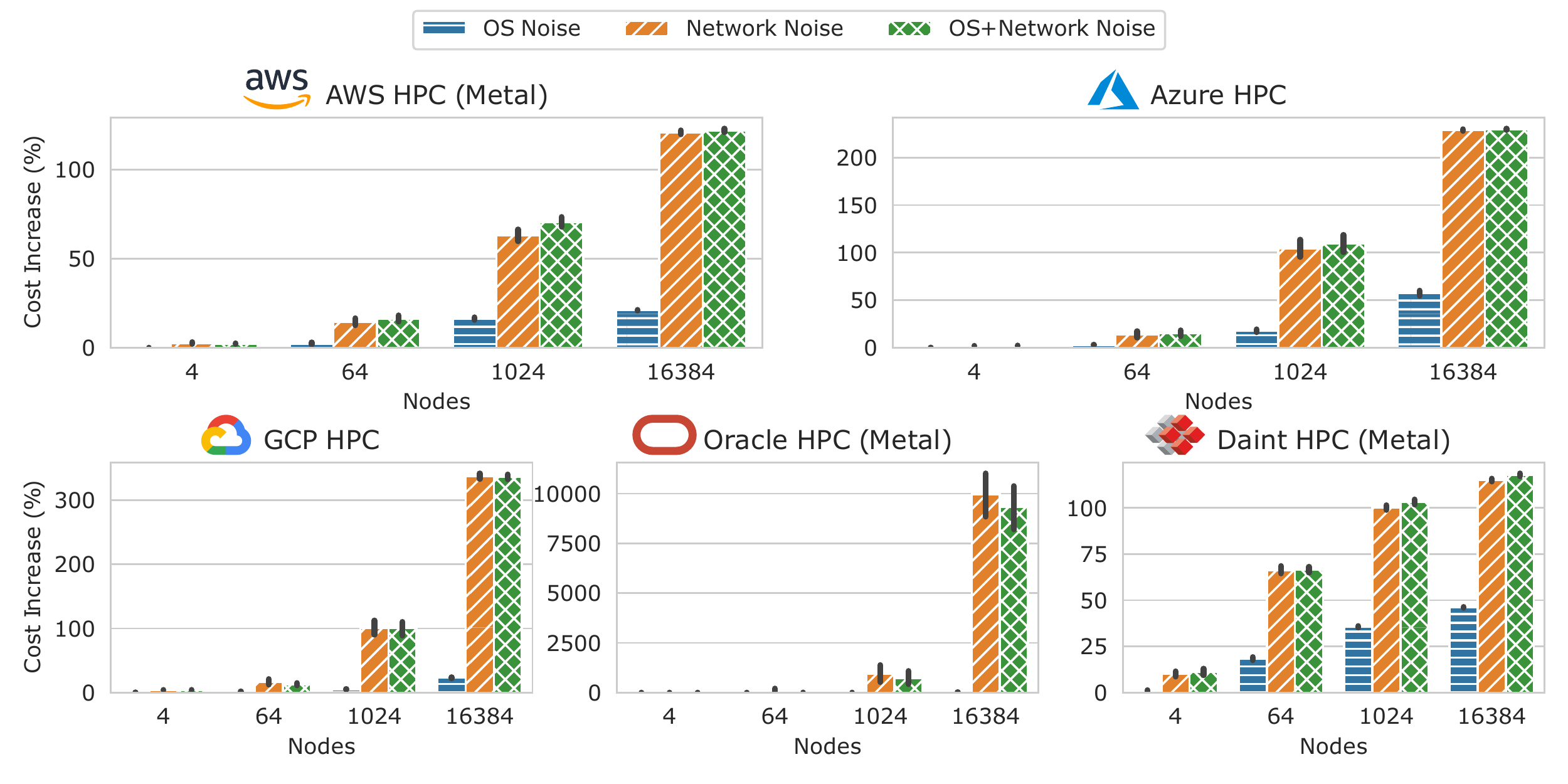}
    \caption{Simulation of the cost $128\times128$ double-precision matrix multiplications followed by a 128KiB dissemination, on different node count. Black vertical lines at the top of the boxes represent the 95\% confidence interval.}
    \label{fig:simulation:cost}
\end{figure*}

We report in Figure~\ref{fig:simulation:cost} the results of this analysis on 100 Gb/s instances for different node count, by showing the relative increase in the monetary cost caused by noise over a noiseless execution, due to the increase in the runtime. We considered the on-demand cost, and we only report the results for those systems where the cost is publicly available (see Table~\ref{tab:systems}). First, we observe that even when running on 64 nodes, the increase in the cost due to noise ranges from 10 to 50\%.

This is exacerbated when increasing the number of nodes, and we observe up to a 2x increase in the cost at 16K nodes. We observe that OS noise only contributes for a fraction of the cost increase, and that much of the increase is caused by network noise. On Oracle, we observe the highest cost increase due to network noise, due to some outliers with very high latency, as discussed in Sec.~\ref{sec:noise:lat} and Sec.~\ref{sec:sim:noise:diss}. Also, it is worth remarking that, a larger cost increase does not necessarily mean a larger cost and that, despite the noise, Azure and Daint are still more cost effective than AWS and GCP for this type of workload.

\begin{figure*}[htpb]
    \centering
    \includegraphics[width=\linewidth]{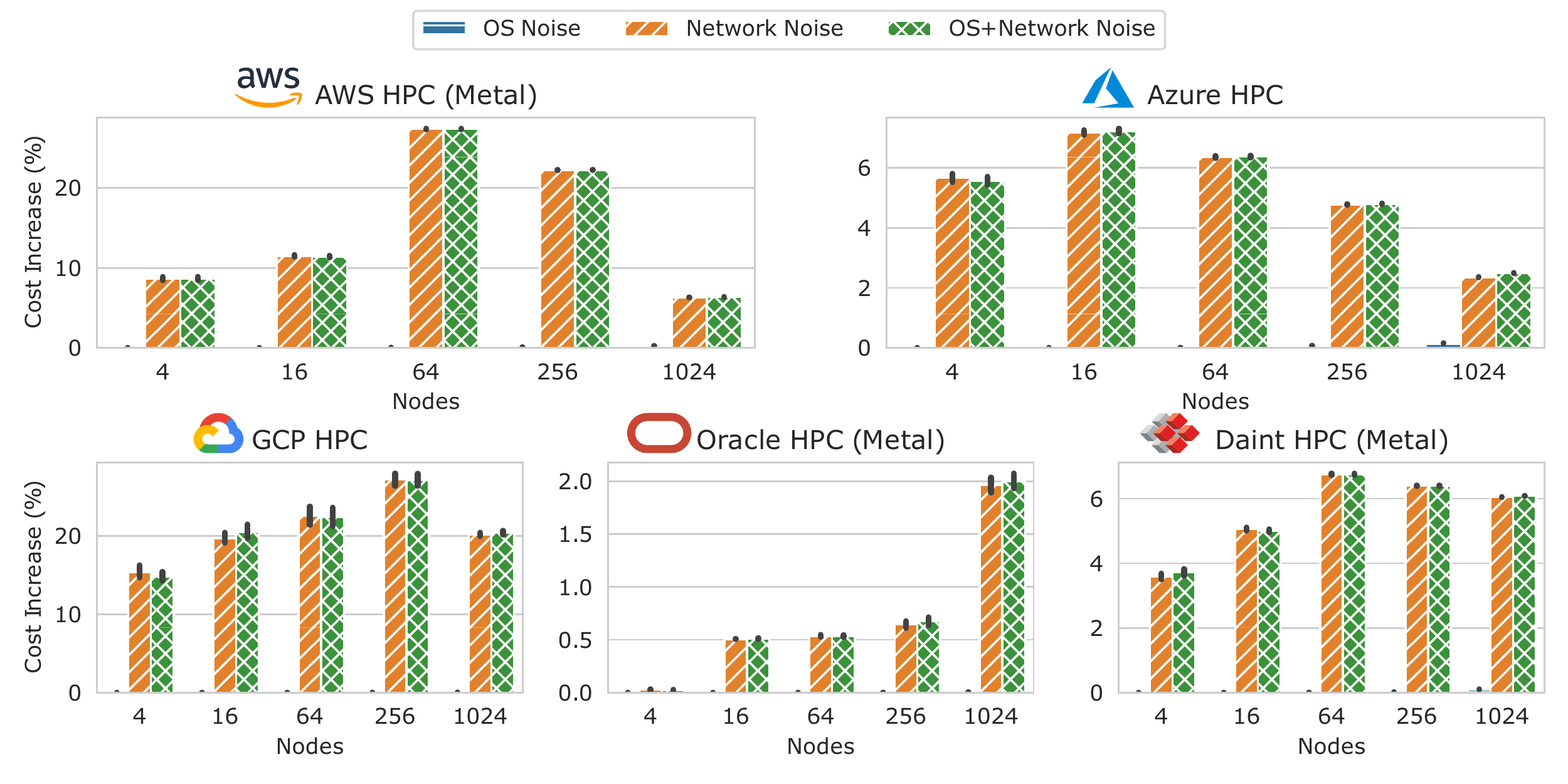}
    \caption{Simulation of the cost $8192\times8192$ double-precision matrix multiplications followed by a 512 MiB ring allreduce, on different node count. Vertical lines at the top of the boxes represent the 95\% confidence interval.}
    \label{fig:simulation:cost_ring}
\end{figure*}

The increase in the cost due to noise also depends on the ratio between communication and computation time. To further analyze this, we repeat the same experiment by simulating matrix multiplication of larger matrices ($8192\times8192$), followed by a 512 MiB allreduce (implemented with a ring collective). While in the previous experiment the communication accounted for more than 90\% of the overall execution time, in this case it accounts for less than 20\% of the overall time. We show the results of this analysis in Figure~\ref{fig:simulation:cost_ring}.

For this workload, OS noise does not play a significant role, since the application is not latency sensitive. On the other hand, network noise increases the monetary cost of at least 5\% on all the systems. On GCP, it causes an additional 15\% cost on 4 nodes, and up to 30\% on 256 nodes.


\begin{shaded*}
\textbf{Observation 10:} Network noise can have a significant impact on the monetary cost of running a distributed application. This is even true at small scale, and for applications that are not dominated by communication.
\end{shaded*}

\section{Discussion}
\subsection{Recommendations}
In general, we observe some differences between the network performance of 100 Gb/s HPC networks of cloud and on-premise HPC systems. For example, AWS and GCP cannot saturate the bandwidth with a single connection. Accordingly, HPC applications running on those two systems should perform, when possible, multiple concurrent communications at any given time to fully exploit the underlying HPC network. Also, both AWS and GCP are characterized by a higher latency than on-premise systems, and are thus not particularly suited for latency-sensitive HPC applications. On the other hand, both Azure and Oracle show latency and bandwidth much similar to those of on-premise HPC systems.

In general, OS and network noise affect in similar ways the performance of both cloud and on-premise HPC systems. However, there are some exceptions. For example, on Oracle we observed some recurrent and severe bandwidth drops (and latency spikes). This also happened on GCP, but only when the two VMs were allocated on different racks. To mitigate the impact of bandwidth noise on GCP, users should take some extra care when creating their VMs, trying to minimize bandwidth-heavy communications between VMs on different racks. On Daint, communications between nodes on different racks also experience latency noise with high intensity. This is partially due to the routing algorithm, and users could use existing techniques to mitigate this issue~\cite{10.1145/3295500.3356196}.

Through simulations we observed that, whereas OS noise has a limited impact on performance, network noise can reduce the performance up to 2x, both at small and large scales. Unfortunately, there is no way for the users to mitigate this, but we want we raised awareness on the impact this issue might have on the performance and cost of running applications (both on cloud and on-premise HPC systems).
\subsection{Limitations}
Our study has some limitations that must be considered to have a complete picture. For example, network noise might depend on which and how many applications were concurrently ran by other users on the systems. Because this is not under our control, we ran each experiment multiple times and at different times of the day, with the aim of capturing different behaviours. However, we observed a consistent behavior. Both Alps and Daint were usually fully utilized, as we observed from Slurm (the job scheduler deployed on those two systems). DEEP-EST was usually quiet, while we could not get any information about how many applications were running at any given time on the cloud systems. 

Similarly, on the cloud we can only control whether the two VMs are on the same or different racks (and not even on all the providers). However, the specific placement and the distance between the VMs would also have a different impact on performance and noise that, however, we are not able to control on such a fine granularity.

Also, it is in general not possible to precisely separate network noise from OS noise. Indeed, part of the network communication is executed on the host and is thus influenced by OS noise. After discussing with Cray HPE engineers, we found out that this is likely one of the reasons why, for example, we observed a high latency noise on Alps. This issue has been fixed through updates to the OS, but those updates were not deployed on Alps before the publication of this paper.

In this work, we used the LogGOPSim simulator, that we extended to simulate latency and bandwidth noise, and instantiated with the parameters we measured on the different systems. Due to limitations in the number of instances we could create on cloud systems, we could validate our simulator only on up to 32 nodes. Although network performance might change at larger scales (e.g., due to a higher average distance between compute nodes, and to higher path diversity), we still believe that these simulations help understanding the general performance trend, and to simulate and isolate the impact of OS and network noise.


\section{Related Work}\label{sec:related}
\subsection{Cloud Benchmarking}
%

%
To improve network performance, recently cloud providers introduced homogeneous HPC instances and low-latency networks with RDMA support. Therefore, various works asked if this could be a game-changer for tightly coupled applications.
Sadooghi et al.~\cite{7045591} compare AWS with a private cloud by considering a wide variety of metrics including memory hierarchy, compute, network, and I/O performance, as well as the cost normalized to memory capacity and bandwidth, and to compute performance. Other metrics include job queueing and turnaround time~\cite{marathe13}. In this work, instead, we focus on network performance and noise, and on its impact on scalability at scale, by performing our analysis on four different cloud providers and three on-premise HPC systems.

Mohammadi and Bazhirov~\cite{mohammadi18} use Linpack benchmarks to compare different cloud computing vendors and a traditional supercomputer. Their results show that Microsoft Azure provides the best results thanks to  the low latency Infiniband interconnect network that facilitates efficient scaling, and performance on the public cloud can be comparable to modern traditional supercomputing systems.
Aljamal et al.~\cite{aljamal20} benchmark various Azure Cloud instances with the NASA parallel benchmark, highlighting the benefits of having high bandwidth capable networking for global communication patterns.
%
%

Guidi et al.~\cite{guidi21}  evaluate benchmarks and applications on AWS, showing that the cloud may have higher bandwidth and lower latency than on-prem systems, especially for medium-large sized messages.
Fernandez~\cite{fernandez21} evaluates single instance and cluster performance on five cloud providers using microbenchmarks to measure performance of collective MPI operations and the HPCG benchmark. Results show that only 100 Gb/s instances exhibits good scalability.

While these works mainly focus on benchmarking of compute performance and application scalability, all of them highlight the importance of the network in achieving profitability. To the best of our knowledge, no other works assess the impact of network noise on cloud. In this paper we provide an in-depth assessment of network performance and noise, and simulate its impact on small collectives at scale. Our simulation methodology can also be applied on other workloads, to understand if and how they can benefit from running on the cloud.

Most of the above works analyze cloud providers using well-known network benchmarks, such as OSU micro-benchmarks~\cite{osu-microbenchmarks} or Intel MPI benchmarks~\cite{imb}. These tools however only provide aggregated measurements over multiple runs, that would then hide noise effects. In this work we rely on the Netgauge measurement tool~\cite{hoefler-netgauge-hpcc07}, that provides detailed per-sample measurements, as well as tools for estimating LogGP parameters and OS noise.

\subsection{Network Noise}
Various works investigate the impact of network noise~\cite{hoefler-collnetnoise, 7877142, groves17}.
Chunduri et al.~\cite{10.1145/3126908.3126926} analyze different sources of performance variability, and the impact of the routing
algorithm on collective operations.
Priscari et al.~\cite{priscari14} propose job allocation strategies to minimize the contention on the links.
Smith et al.~\cite{8665797} study the impact of network noise on both
dragonfly and fat-tree networks and propose an adaptive routing
algorithm for fat-trees. De Sensi et al.~\cite{10.1145/3295500.3356196} propose an application-aware routing algorithm to mitigate network noise and improve application performance on dragonfly networks. However, all these works focus on on-prem systems and, to the best of our knowledge, this is the first systematic analysis of network noise on HPC cloud systems and on its impact on scalability. 





\section{Conclusions}\label{sec:conclusions}


The HPC community often looks at cloud systems with skepticism, amongst the concerns about their costs, and the suspicion about their ability to
efficiently run large-scale tightly coupled applications. In this work, we  analyze network performance and noise, and their impact on the scalability of large-scale computations. 

Like on-prem systems, cloud interconnection networks suffer from network performance variability. 
We used small scale performance and noise measurements to assess the impact of noise at a larger scale through simulations.
After validating the simulation environment, we showed that all the providers are affected by OS and network noise both at small and large scale, and for different communication patterns. Those effects must be taken into account when we look at the cost profitability of such systems.

Although cloud systems are updated frequently,
we believe the methodology we used (benchmark at a small scale and simulation at a larger scale) can be utilized on any large scale system to evaluate the impact of OS and network noise at scale. This approach can be a valuable tool for administrators and architects to assess the extent of OS and network noise. Also, it can help researchers, institutions, and businesses that want to quickly evaluate the potential impact of noise on the scalability of a cloud based solution for their HPC workloads without paying the cost of renting (and getting access to) a large number of machines. 

\section*{Acknowledgment}
We would like to thank for their support, feedback, and insightful discussions: CSCS; Duncan Roweth from Cray HPE; Evan Burness from Microsoft Azure; Matt Koop and Brendan Bouffler from AWS; Bill Magro, David Wetherall, Jiuxing Liu, and Rick Jones from GCP; Patrick Saltzmann, Anupam Karmakar, and Calebe Kuenzle from Oracle Cloud. We also thank our shepherd, Kevin Vermeulen, and all anonymous reviewers for their insightful feedback and suggestions, which substantially improved the content and presentation of this paper. 

This work has been partially funded by the European Research Council (ERC) \includegraphics[height=1.3em]{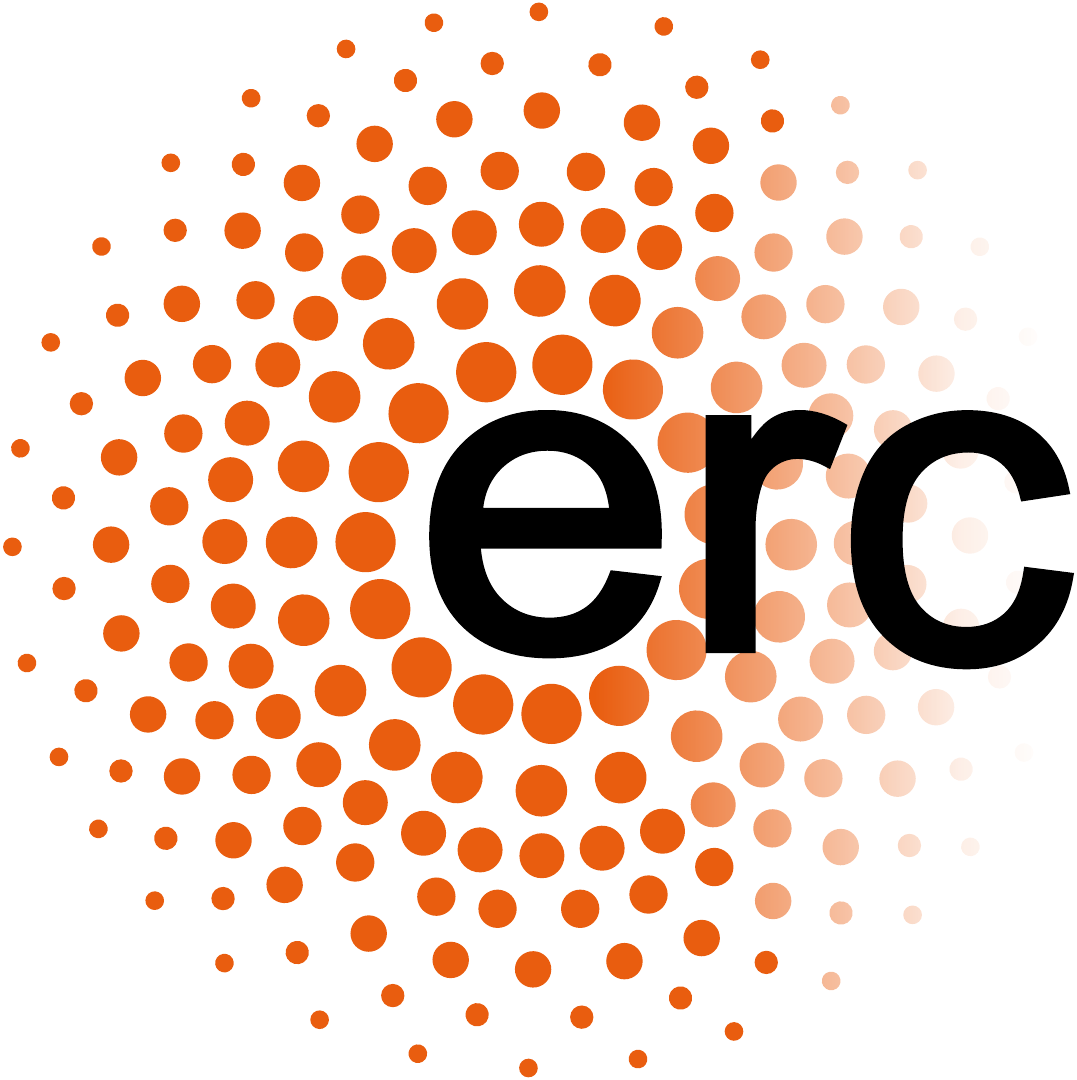} grant PSAP (grant no. 101002047), the European Union’s Horizon Europe programme projects DEEP-SEA (grant no. 955606), and RED-SEA (grant no. 955776). Daniele De Sensi was supported by an ETH Postdoctoral Fellowship (19-2 FEL-50).

\bibliographystyle{IEEEtran}
\bibliography{main}

\appendices
\begin{figure*}[htpb]
    \centering
    \includegraphics[width=\linewidth]{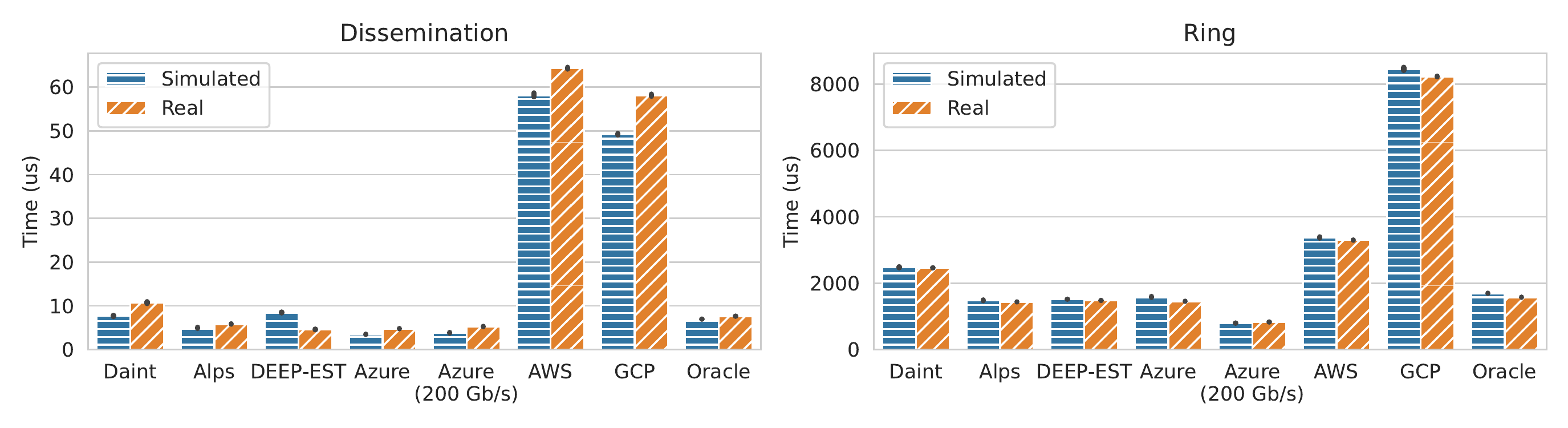}
    \caption{Comparison between measured (on HPC instances) and simulated times for 16B on dissemination and 16MiB ring collectives on 4 nodes. Black vertical lines at the top of the boxes represent the 95\% confidence interval.}
    \label{fig:simulation:validation:4}
\end{figure*}

\begin{figure*}[htpb]
    \centering
    \includegraphics[width=\linewidth]{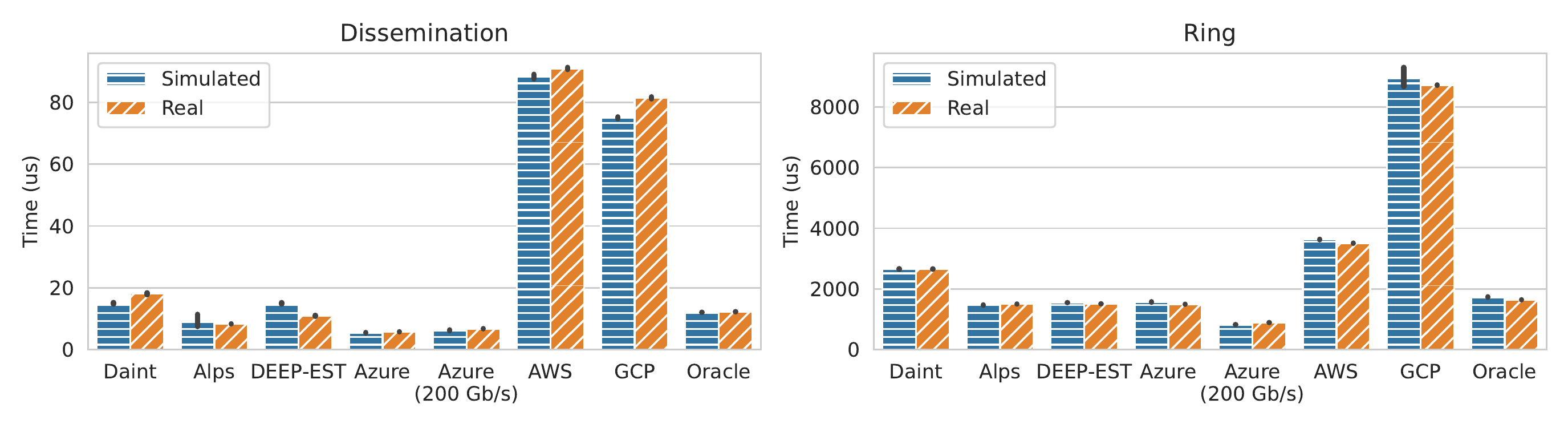}
    \caption{Comparison between measured (on HPC instances) and simulated times for 16B on dissemination and 16MiB ring collectives on 8 nodes. Black vertical lines at the top of the boxes represent the 95\% confidence interval.}
    \label{fig:simulation:validation:8}
\end{figure*}

\begin{figure*}[htpb]
    \centering
    \includegraphics[width=\linewidth]{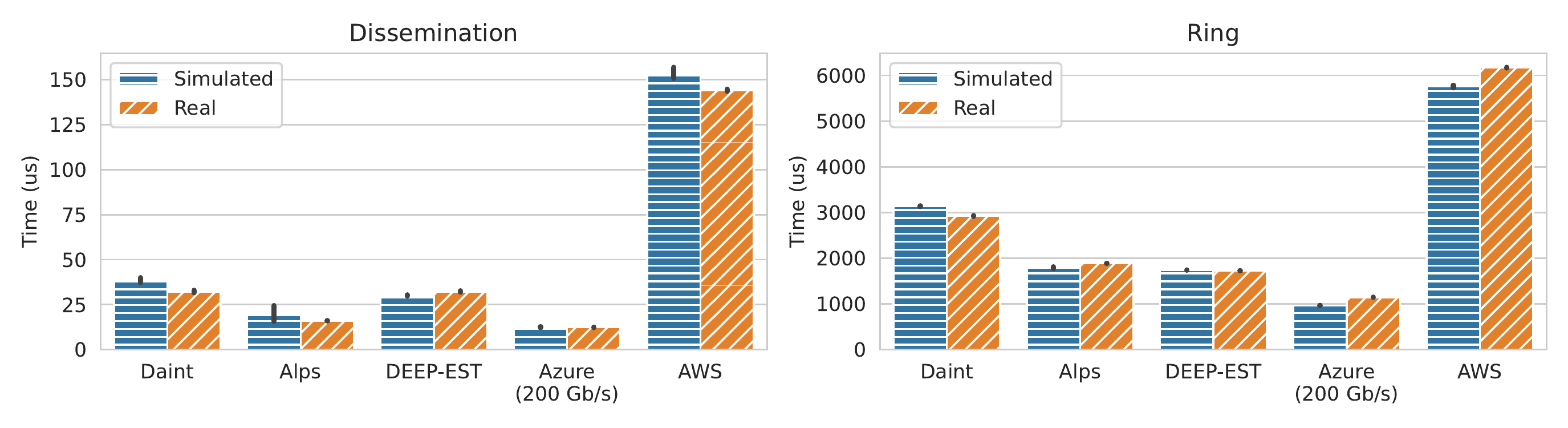}
    \caption{Comparison between measured (on HPC instances) and simulated times for 16B on dissemination and 16MiB ring collectives on 32 nodes. Black vertical lines at the top of the boxes represent the 95\% confidence interval.}
    \label{fig:simulation:validation:32}
\end{figure*}
\section{Simulator Validation}\label{sec:appendix:validation}
We repeated the validation process described in Sec.~\ref{sec:simulations:validation} on 4 and 8 nodes and, for the instance types where this was possible, also on 32 nodes (not all the providers sufficiently increased our quota limits to enable the creation of a cluster with 32 nodes). We report the results for 4, 8, and 32 nodes in Figure~\ref{fig:simulation:validation:4}, Figure~\ref{fig:simulation:validation:8}, and Figure~\ref{fig:simulation:validation:32} respectively. Similarly to the results shown in Sec.~\ref{sec:simulations:validation}, we observe that the simulated runtime closely matches the measured one.

\section{Clusters Provisioning Time}
A cluster composed of 2 HPC instances can be provisioned in 22 minutes on Azure, AWS, and Oracle. On the other hand, on GCP this requires less than 1 minute. However, it is worth remarking that on Azure, AWS, and Oracle, the cluster is already provisioned with Slurm and with many software packages often needed for running HPC applications (MPI, Intel MKL, etc...). On the other hand, on GCP everything needs to be installed and configured from scratch and, eventually, the time required to have a fully provisioned cluster is similar on all the four cloud providers.

\end{document}